\documentclass[12pt,preprint]{aastex}
\newcommand{\kms}{{\,\rm km\,s}^{-1}}

\renewcommand{\mag}{\mbox{$\;$mag}}
\begin{document}
% ******************************************************************

\title{\boldmath
     The Linearity of the Cosmic Expansion Field from 300 to
     $30,000\kms$ and the Bulk Motion of the Local Supercluster with
     Respect to the CMB}

\author{A. Sandage}
\affil{%
     The Observatories of the Carnegie Institution of Washington,\\
     813 Santa Barbara Street, Pasadena CA, 91101, USA}

\author{B. Reindl and G.~A. Tammann}
\affil{%
     Department of Physics and Astronomy, Univ.\ of Basel, \\
     Klingelbergstrasse 82, 4056 Basel, Switzerland}
\email{g-a.tammann@unibas.ch}

% ******************************************************************
%    Abstract
% ******************************************************************
\begin{abstract}
The meaning of ``linear expansion'' is explained. Particularly
accurate {\em relative\/} distances are compiled and homogenized a)
for 246 SNe\,Ia and 35 clusters with $v<30,000\kms$, and b) for
relatively nearby galaxies with 176 TRGB and 30 Cepheid distances. 
The 487 objects define a tight Hubble diagram from $300-30,000\kms$
implying individual distance errors of $\la7.5\%$.
Here the velocities are corrected for Virgocentric
steaming (locally $220\kms$) and -- if $v_{220}>3500\kms$ -- for a
$495\kms$ motion of the Local Supercluster towards the warm CMB pole
at $l=275$, $b=12$; local peculiar motions are averaged out by large
numbers. A test for linear expansion shows that the corrected
velocities increase with distance as predicted by a standard model 
with $q_{0}=-0.55$ [corresponding to 
$(\Omega_{\rm M},\Omega_{\Lambda})=(0.3,0.7)$],  
but the same holds -- due to the distance limitation of the present
sample -- for a range of models with $q_{0}$ between $\sim\!0.00$ and
$-1.00$. For these models $H_{0}$ does not vary systematically by more
than $\pm2.3\%$ over the entire range. Local, distance-dependent
variations are equally limited to 2.3\% on average. In particular the
proposed Hubble Bubble of Zehavi et~al. and Jha et~al. is rejected at
the $4\sigma$ level. --
Velocity residuals in function of the angle from the CMB pole yield 
a satisfactory apex velocity of $448\pm73\kms$ and a coherence radius
of the Local Supercluster of $\sim3500\kms$ ($\sim56\;$Mpc),
beyond which galaxies are seen on average at rest in co-moving
coordinates with respect to the CMB. Since no obvious single
accelerator of the Local Supercluster exists in the direction of the
CMB dipole its motion must be due to the integral gravitational force
of all surrounding structures. Most of the gravitational dipole comes
probably from within $5000\kms$.
\end{abstract}
% ******************************************************************

% ******************************************************************
% 1. Introduction
% ******************************************************************
\section{\MakeUppercase{Introduction}}
\label{sec:1}
%
% ******************************************************************
% 1.1. Preliminaries
% ******************************************************************
\subsection{Preliminaries}
\label{sec:1:1}
The basic prediction of all models of ideal universes that are
homogeneous and isotropic is that, if the expansion is real, there
must necessarily be a linear relation between redshift and
distance. There are many proofs, but among the earliest are those by
\citet{Lemaitre:27,Lemaitre:31} and \citet{Robertson:28} even before
the observational announcement by \citet{Hubble:29} in which he, 
in the final sentence of his paper, suggested linearity may be only
local. 

     Modern theoretical proofs for global linearity are set out in
many of the standard text books at various levels of mathematical
sophistication. Popular among the simpler proofs is that a linear
relation is the only one that preserves relative shapes of geometrical
shapes in an expanding manifold, and which is the same from all
vantage points. Deeper proofs, based on properties of the metric of
homogeneous, isotropic models, follow
\citet{Robertson:29,Robertson:33} and \citet{Walker:36}, and lead to
the Robertson-Walker line element of the metric.   

     A unique property of a linear velocity field is that every
vantage point in the field appears to be the center of the expansion
and has the same ratio of velocity to distance over the entire field.
Because of this important property, much effort over the past 80 years
has been made by the observers to prove, or disprove, linearity,
either locally or globally. The first comprehensive result was that by 
\citet{Hubble:Humason:31}, enlarging the observational data available
to Hubble in \citeyear{Hubble:29}.  

     Many summaries of the theoretical expectation of linearity and
observational verification for the standard model exist. From the
theoretical side the classic text books include those by 
\citet{Heckmann:42}, 
\citet{Bondi:60},
\citet{Robertson:Noonan:68},
\citet{Harrison:81},
\citet{Narlikar:83},
\citet{Peebles:93},
\citet{Peacock:99}.
\citet{Bowers:Deeming:84}, and \citet{Carroll:Ostlie:96} are exemplary
at the high end of the intermediate level. On the observational side,
text book-like chapters by \citet{Gunn:Oke:75}  
and \citet{Sandage:75,Sandage:88,Sandage:95} are useful.  

     Deviations from a linear velocity-distance relation divide into
two categories. 
(1) The relation is non-linear everywhere, i.e.\ the deviation is global.
(2) The deviation is local, going over into a linear relation at large
distances.

     The first category is the most fundamental because it denies the
standard homogeneous model everywhere. The second contains models
where either the local Hubble constant changes up to a certain
distance such as increasing or decreasing outward, or where the local
irregularities are due to streaming motions due, presumably, to a
local inhomogeneous distribution of matter. 

     The most radical of the first category of global non-linear
models can easily be disproved observationally, even as early as in
\citet{Hubble:Humason:31}, by noting that the slope of the local
log(redshift) - magnitude relation is close to 0.2 as expected in the
case of linear expansion. 
This conclusion was much strengthened by the tight Hubble diagram
from brightest clusters galaxies in \citet{Humason:etal:56} and then,
from 1972 to 1975, by a series of papers on the velocity-distance
relation based on new redshifts and apparent magnitudes of
galaxy clusters measured at Mount Wilson and Palomar 
\citep[][Table~1 for a summary]{Sandage:99}.   

     Non-linear models in the second category are more difficult to
disprove because the deviations from the pure Hubble linear flow are
much smaller than in the radical first group and can easily be mocked
by systematic distance errors due to bias effects. 
The literature is large on deviations from a pure Hubble flow, either
due to local streaming motions or to a variation of the Hubble
constant outward, or to a local bulk motion relative to a globally
significant distant kinematic frame. Some are easier to disprove than
others.  

     One such non-linear model assumes that the local global
distribution of matter is hierarchical. Following 
\citet{Charlier:08,Charlier:22} and using a suggestion by
\citet{Carpenter:38} of a density-size relation for all objects in the
universe, \citet{deVaucouleurs:70,deVaucouleurs:71} postulated a
universe made of hierarchies of decreasing mean density 
with increasing volume size up to some limiting distance.
The velocity-distance relation in such a universe was formulated by 
\citet{Haggerty:70}, and \citet{Haggerty:Wertz:71}, following a
prediction of \citet{Wertz:70}, and was found to be nearly quadratic
locally, but becoming linear at large distances. The model was shown,
however, to be irreconcilable with observations  
\citep{STH:72}.

     It was early demonstrated \citep[e.g.][]{%
ST:75,Teerikorpi:75a,Teerikorpi:75b,Teerikorpi:84,Teerikorpi:87,
Fall:Jones:76,Bottinelli:etal:86,Bottinelli:etal:87,Bottinelli:etal:88, 
Sandage:88,Sandage:94,FST:94}
that the claimed changes of the Hubble constant outward were 
often caused by uncorrected observational selection bias 
\citep[see also][Chapter~10]{Sandage:94,Sandage:95}.%
% ******************************************************************
% Footnote 1
% ******************************************************************
\footnote{It is important to emphasize that bias of this nature is 
   {\em not\/} the famous Malmquist bias, but rather is an
   incompleteness bias that increases with distance. In contrast, the
   Malmquist bias is distance independent. It only concerns the error
   made in assigning a number to an absolute magnitude calibration of
   a distance indicator when using a sample that is magnitude, rather
   than, distance limited. On the other hand, an incompleteness bias
   causes an error that increases with distance, which, if
   uncorrected, makes the Hubble constant appear to increase outward
   with distance. The error of calling the incompleteness bias as
   Malmquist is wide spread in the literature.}
% ******************************************************************
It must also be stated that in the early papers some of the claimed
distortions of the velocity field were due to the omission of velocity
corrections for Virgocentric infall and -- for the more distant
galaxies -- for CMB dipole motion.

     Hence, despite many papers and conferences on proposed streaming
motions and searches for variations of the Hubble constant with 
distance, the problem is still open. What has been needed are
distance indicators that have very small intrinsic dispersion so as to
eliminate, or greatly reduce, the effect of distance-dependent
incompleteness selection bias. 
At small distances the increasing number of Cepheid and TRGB distances 
becomes most helpful here, whereas for large distances 
Type Ia supernovae at maximum light have emerged as the best standard 
candles, just as \citet{Zwicky:62} had proposed.

% ******************************************************************
% 1.2. The complications: the invisible world map must be
%      inferred from the visible world picture
% ******************************************************************
\subsection{The Complications: The Invisible World Map
            Must be Inferred from the Visible World Picture}
\label{sec:1:2}
The concept of linearity of a velocity-distance velocity field, so
simple to describe, is most complicated to define properly in an
expanding universe. \citet{Milne:35} proposed language that clarifies
the problem. \citet{Robertson:55} also uses the same language. 

     The problems are these.

      (1) Because we do not observe the universe at large 
redshift at the same cosmic time as at low redshift, we must 
distinguish between what Milne called the world map and the 
world picture. The world map is the state of the universe at a      
given cosmic time. The world picture is what appears to us. We 
cut the series of world maps at different cosmic times as we look 
to different redshifts. 

     By a linear ``velocity''-``distance'' relation is meant that at a
particular cosmic time there is a linear relation in the world map
between the coordinate distance as measured by the metric, and
redshift. But the metric distance, $R(t)r$, where $R(t)$ is the time
dependent expansion scale factor and $r$ is the invariant (constant
for all time) comoving radial metric ``distance''. It is not the same
as a distance measured by an astronomer using observational data.
There are many kinds of such ``astronomical'' distances, depending on
the method by which they are measured. There is the distance when
light left a galaxy, the distance when light is received, the
luminosity distance by apparent magnitude, the distance by angular
size, the round-trip distance by radar signals. All differ at large
redshifts, and are the same only in the zero redshift limit. 
\citet*{McVittie:74} discussion is useful here. 
Which distance do we use in formulating a velocity-``distance''
relation and proving that it is ``linear''?     

     There is also the difficulty with ``velocity''. What we measure
is redshift, not velocity. The two are not the same except, again, in
the zero redshift limit \citep{Harrison:93}.  

     The problem is solved by not using the complicated concepts of
velocity and distance but by transforming the world map into the
observer's world picture {\em using only observables}. Before 1958
this mapping was done by making a Taylor series expansion of the
$R(t)$ expansion factor about the present cosmic time so as to sample
the world map at the earlier cosmic times \citep{Robertson:55}. 
This could be called a tangent mapping, and is not very useful for
redshifts larger than those at moderately local distances. 
What was needed was a general transformation mapping that is valid for
all redshifts. 

     \citet{Mattig:58} solved the problem with his famous equation
that relates the metric distance of a galaxy, $R_{0}r$, at the time of
light reception with the redshift, no matter how large, in models with
zero cosmological constant. Metric distance is further replaced by
apparent magnitude by the \citet{Robertson:38} equation that connects
the two. In addition to the apparent magnitude and the redshift the
equation contains the \citet{Robertson:55} deceleration parameter,
$q_{0}=-R_{0}\ddot{R}/\dot{R}^{2}$, which, in the simplest models with
no cosmological constant, is determined by the mean matter density of
the model. In that case the values of $H_{0}$ and $q_{0}$ observed at
the present epoch are used to parametrize the past and the future of
the world map. 

     The Mattig equation, or more complicated versions for models 
that incorporate the cosmological constant
\citep[e.g.][]{Mattig:59,Carroll:etal:92} which we use in the next
section, provides the  connection between the world map and the world
picture. It has properly been said that the Mattig equation is 
``one of the single most useful equations in cosmology as far as
observers are concerned'' \citep[][p.~89]{Peacock:99}. Mattig's
solution began the modern era of practical observational cosmology and
has been used for many auxiliary and related problems concerning such
observational data as galaxy counts, angular diameters, and others
throughout the past 50 years. 

     Because the world map defined in the Robertson-Walker metric has
a well defined linear relation between coordinate (metric) distance
and redshift, and because a Mattig-like equation transforms the map in
to the observed picture, the test for linearity becomes one of
comparing the observed redshifts at a given apparent magnitude with
the predicted redshifts from a model. The linearity test becomes,
then, a search for residuals in the subtraction of the redshifts
predicted {\em from the model\/} from the observed redshifts at a
given photometric distance obtained from the observations as
$(m-M)$. Streaming motions and/or variations of the Hubble constant
outward are searched for as correlations (or not) of the residuals
with direction and photometric distance.

     The test for linearity is, then, model-dependent: we must compare
the observed redshift-magnitude relation with the prediction for some
adopted world model as it is transformed into the world picture by a
Mattig-like equation of $(m-M)=f(z,q_{0},\Lambda)$. We can only test
linearity relative to the adopted model. We test for the sensitivity
to the adopted model in \S~\ref{sec:3}.

% ******************************************************************
% 1.3. This paper
% ******************************************************************
\subsection{This Paper}
\label{sec:1:3}
The purpose of this paper is:
(1) To test the linearity of the local expansion field within 
$z<0.1$. Beyond this limit the linearity of the expansion of
space is well documented out to $z=0.4$ by the SNe\,Ia of
\citet{Hicken:etal:09}, which have ample overlap with the present
data, and which can be tightly fit to the SNe\,Ia compiled by 
\citet{Kessler:etal:09} extending to $z>1$. The high-$z$ data are
essential to optimize the determination of the world model for which
linear expansion is valid. But at smaller distances the
character of the expansion field is still poorly known because so far
the number of objects with sufficiently accurate distances has been
small. Yet the test is important to understand the effect of the
observed clumping of visible matter on the local dynamics. It is also
important to find the minimum distance at which the cosmic value of
$H_{0}$ can be found independent of peculiar velocities. Strong local
deviations from linearity have in fact frequently been claimed up to
the recent past by \citet{Zehavi:etal:98} and \citet{Riess:etal:09}. 
(2) To determine the size of the comoving volume that partakes of
our observed velocity relative to the dipole of the CMB.

     The plan of the paper is this. In \S~\ref{sec:2} the data are
specified. In \S~\ref{sec:2:1} accurate {\em relative\/} distances of
246 SNe\,Ia and 35 clusters within the adopted distance range are
compiled from eight different sources. The distances have demonstrably
rms errors of $\le0.18\mag$; only relative distances are needed in the
present context. The distances on an arbitrary zero point, reduced to
the barycenter of the Local Group, are listed in Table~\ref{tab:ml}
together with the appropriate velocities $v_{\rm hel}$, $v_{220}$ 
(corrected for Virgocentric infall), and $v_{\rm CMB}$ (corrected for
motion with respect of the CMB dipole $A_{\rm corr}$ if 
$v_{220}>3500\kms$). 
In \S~\ref{sec:2:2} 30 Cepheid and 176 TRGB distances which are
equally accurate as those of SNe\,Ia are added in order to extend the
sample down to $300\kms$. They are reduced to the barycenter of the
Local Group and corrected for the Virgocentric flow as in
\S~\ref{sec:2:1}. We justify in \S~\ref{sec:2:3} why other distance
indicators are not considered here.

     In \S~\ref{sec:3} the problem of linear expansion is set out. 
The actual test in \S~\ref{sec:3:1} uses a Hubble diagram of 
{\em all\/} objects in the sample and analyzes the residuals from a
Hubble line for a standard $\Lambda$CDM model with  
$\Omega_{\rm M}=0.3$ and $\Omega_{\Lambda}=0.7$. 
(Note that this model implies a deceleration parameter of
$q_{0}=-0.55$ because it holds that 
$q_{0}=0.5(\Omega_{\rm M}-2\Omega_{\Lambda})$ for all Friedmann models
[e.g.\ \citealt{Sahni:Starobinsky:00}]). \S~\ref{sec:3:2} explores the 
dependence of linearity on the adopted model in terms of $q_{0}$.

     An analysis of the velocity residuals of objects with 
$v_{220}<7000\kms$ (the limit is set to avoid an overwhelming effect
of distance errors) in function of direction is in \S~\ref{sec:4},
showing the size of the Local Supercluster and its motion toward
$A_{\rm corr}$, which is the direction of the warm pole of the
CMB after correction for our Virgocentric velocity vector.

        Results and conclusions are in \S~\ref{sec:5}.

% ******************************************************************
% 2. The data
% ******************************************************************
\section{\MakeUppercase{The Data}}
\label{sec:2}
In order to trace the local expansion field the 
{\em most accurate relative\/} distances are compiled of objects,
comprising SNe\,Ia, clusters, Cepheids, and tip of the red-giant
branch (TRGB) distances, with velocities from $300$ to $30,000\kms$.

     All quoted distances are reduced to the barycenter of the 
Local Group (LG) which is assumed to lie on the line between the
Galaxy and M31 and at a distance of one third of $(m-M)_{\rm M31}$.
The small random error of the distance moduli ($0.15\mag$ on
average as shown below) makes the samples unusually insensitive to
incompleteness bias. 

     Heliocentric velocities $v_{\rm hel}$ are available for all
objects. They are reduced here to the barycenter of the LG following
\citet{Yahil:etal:77}. The resulting velocities, $v_{\rm LG}$, are
then corrected for a self-consistent Virgocentric infall model by
$\Delta v_{220}$ which is the vector sum of the 
$220 (\pm\sim\!30)\kms$ infall vector at the position of the LG,  
\citep{Peebles:76,Peebles:80,Yahil:etal:80a,Hoffman:etal:80,
Tonry:Davis:81,Dressler:84,Yahil:85,Tammann:Sandage:85,Kraan-Korteweg:86,
deFreitasPacheco:86,Giraud:90,Jerjen:Tammann:93}  
and the infall velocity of a particular object, both projected onto
the line of sight between the observer and the object.
The center of the Virgo cluster is taken to coincide with NGC\,4486 at
$l=283.8$, $b=74.5$.
The infall vectors scale with Virgocentric distance $r_{\rm Virgo}$ 
like $1/r_{\rm  Virgo}$ if a density profile of the Local Supercluster of 
$\rho\sim r^{-2}$ is assumed \citep{Yahil:etal:80b}. Since good (relative)
distances are known for all objects including the Virgo cluster, the
corrections  $\Delta v_{220}$ can be calculated from 
equation~(5) in \citet{STS:06}. 
The velocities $v_{220}=v_{\rm LG}-\Delta v_{220}$ would be 
observed in the absence of any streaming towards the Virgo cluster
center and if the infall model were exact, but deviation from the
model \citep[see e.g.][]{Klypin:etal:03} will add to the true peculiar
velocities of local galaxies. The velocities $v_{\rm CMB}$ of only
the objects with $v_{220}>3500\kms$ are in addition corrected by 
$\Delta v_{\rm CMB}=v_{220}-495\cos\alpha\kms$
($\alpha$ being the angle from the CMB apex $A_{\rm corr}$ 
to compensate for the reflex of the Local Supercluster motion relative
to the CMB [see \S~\ref{sec:4}]).

% ******************************************************************
% 2.1 SN Ia and cluster distances
% ******************************************************************
\subsection{SN\,Ia and Cluster Distances}
\label{sec:2:1}
Eight published sets of accurate relative distances that contain more
than 10 SNe\,Ia or clusters are considered here.  

     Six sets are based on SNe\,Ia. Since SNe\,Ia are treated as
standard candles it makes no difference whether their distance
measures are published as corrected apparent magnitudes at maximum or
as distance moduli. In either case the data are converted to ``true''
distance moduli by forcing each data set to give the same fixed mean
value of $H_{0}$. We have chosen here 
$H_{0}=62.3$ [km$\;$s$^{-1}\;$Mpc$^{-1}$] as found from
Cepheid-calibrated SNe\,Ia \citep{STS:06,TSR:08b}. 
As mentioned before the absolute calibration is not necessary for the
present investigation, but a realistic calibration will simplify some
of the following discussions.

     Two sets of relative cluster distances expressed in $\kms$ yield
equally tight Hubble diagrams as SNe\,Ia. This is thanks to the large
number of galaxy distances determined in each cluster. The mean
cluster distances are normalized to the same fiducial value of $H_{0}$.

\begin{figure}[t] % [p]
   \epsscale{0.8}
\plotone{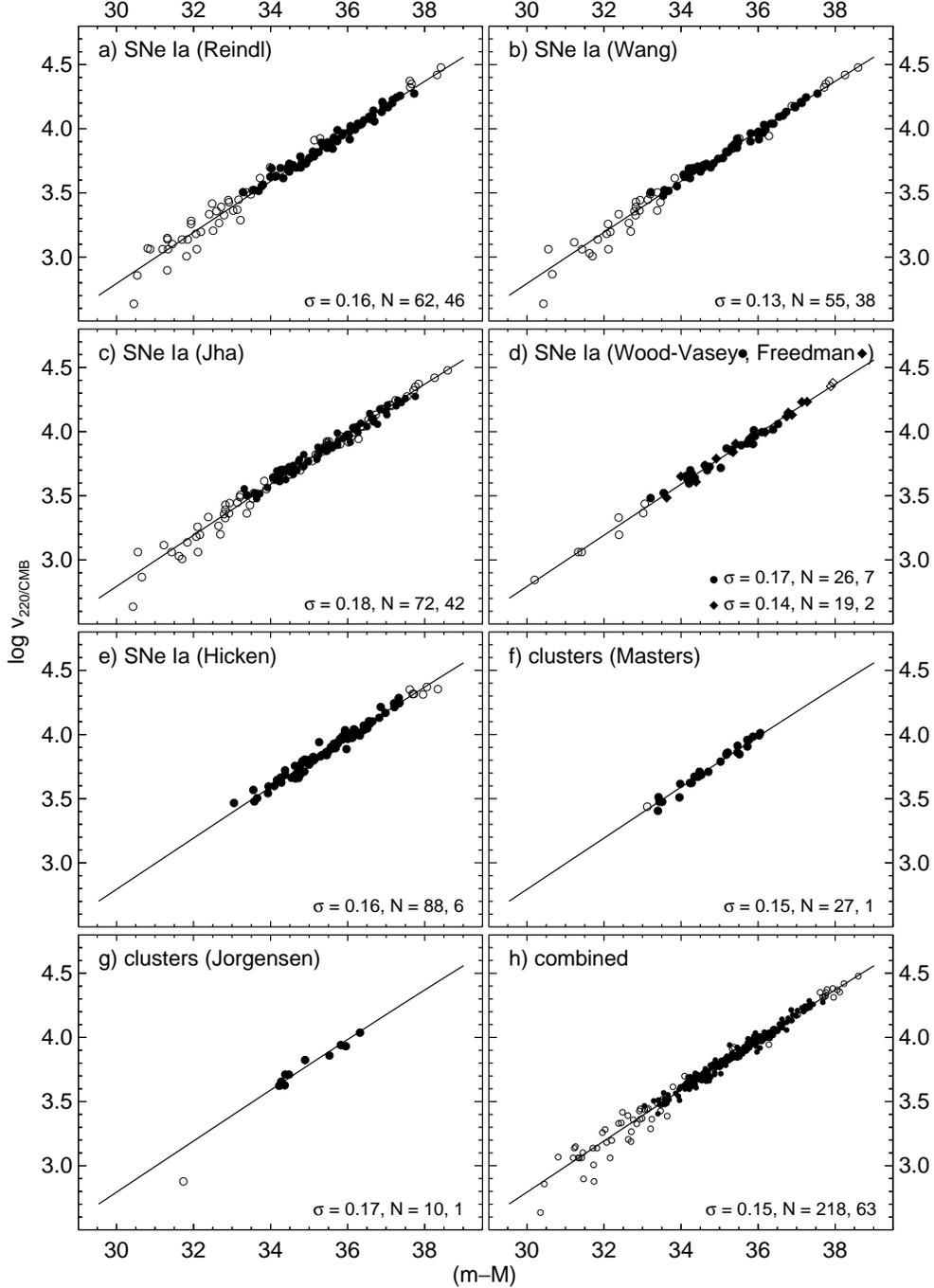}
  \caption{a)$-$g) Hubble diagrams from eight different sources shifted
    to a common value of $H_{0}$ (Here $H_{0}=62.3$). The two values
    of $N$ give the number of objects with 
    $3000<v_{220/{\rm CMB}}<20,000\kms$ (closed symbols) and those
    outside this range (open symbols). Two SNe\,Ia with $(m-M)<30.0$
    are not shown. The solid slightly curved line, defined by 
    equation~(\ref{eq:carroll}), holds for a flat
    $\Omega_{\rm M}=0.3$, $\Omega_{\Lambda}=0.7$ model. 
    h) The combined Hubble diagram from a)--g); objects with more
    than one distance determination are plotted at the mean
    distance. The increasing scatter below $(m-M)\sim33.0$ is a clear
      effect of the relative importance of peculiar velocities at
      small redshifts.} 
  \label{fig:01}
\end{figure}

     The objects in each data set are separately plotted in Hubble
diagrams $\log v_{\rm 220/{\rm CMB}}$ vs.\ $(m-M)$ in
Figure~\ref{fig:01}a-g. 
The Hubble line shown holds for the adopted $\Lambda$CDM model with
$\Omega_{\rm M}=0.3$, $\Omega_{\Lambda}=0.7$ and is defined by
\citep{Carroll:etal:92}
\begin{equation}\label{eq:carroll}
(m-M) = 5\log c(1+z_{1})\int_{0}^{z_{1}}[(1+z)^{2}(1+\Omega_{\rm
  M}z)-\Omega_{\Lambda}z(2+z)]^{-1/2}dz + 25 -5\log H_{0}.
\end{equation}
The scatter about the Hubble line ($0.13-0.18\mag$) is shown in each
panel. The values refer to the velocity range of
$3000<v_{220/{\rm CMB}}<20,000\kms$ where they are least affected by
peculiar motions, $K$-corrections, and photometry at faint levels.

% ********************************************************
% Figure 1: Hubble diagrams of (1) - (8)
% ********************************************************

     The sources of the eight samples -- all normalized here to the
fiducial value of $H_{0}$ -- are detailed in the following:

     (1) Maximum magnitudes $m^{\rm corr}_{V}$ are available for 105
Type Ia supernovae (SN\,Ia) (62 of which fall into the fiducial range
$3000<v_{220/{\rm CMB}}<20,000\kms$) from \citet{Reindl:etal:05},
excluding spectroscopically peculiar objects of type SN\,1991T and
1991bg. They have been corrected for Galactic and internal absorption
and homogenized as to decline rate and intrinsic color. The
corresponding distances yield the Hubble diagram in
Figure~\ref{fig:01}a.

     (2) \citet{Wang:etal:06} give corrected distance moduli based on
maximum $UBVI$ magnitudes and some adopted standard luminosity of 98
SNe\,Ia (55 of which fall into the fiducial velocity range). The
magnitudes are corrected for absorption and homogenized as to decline
rate and color index with somewhat different precepts as under (1).
Their published SN magnitudes are brighter than in (1) because the
intrinsic color of SNe\,Ia was assumed to be improbably blue causing
large absorption corrections, but this does not disturb the internal
consistency of the data. The data give the Hubble diagram in
Figure~\ref{fig:01}b.

     (3) \citet{Jha:etal:07} have derived homogenized,
absorption-corrected maximum $V$ magnitudes of 95 SNe\,Ia (72 of which
fall into the fiducial velocity range) by fitting multi-color light
curves to templates. The SNe\,Ia are plotted in Figure~\ref{fig:01}c.

     (4) Maximum $H$  magnitudes of 33 SNe\,Ia (of which 26 fall into
the fiducial velocity range) have been published by
\citet{Wood-Vasey:etal:08}. They provide, {\em without\/} any
homogenization for decline rate or color and with only insignificant
absorption corrections, the Hubble diagram shown in
Figure~\ref{fig:01}d. 

     (5) Maximum $I$ magnitudes, corrected for absorption and decline
rate, for 21 SNe\,Ia (of which 19 fall into the fiducial velocity range)
have been published by \citet{Freedman:etal:09}. The data have been
added in Figure~\ref{fig:01}d.

     (6) \citet{Hicken:etal:09} have reduced SN data for a very large
sample of SNe\,Ia in four different ways. Two methods follow 
\citet{Guy:etal:05,Guy:etal:07} and two methods follow the multi-color
light curve fitting of \citet{Jha:etal:07} with two different
assumptions on the absorption-to-reddening ratio for SNe\,Ia
($R_{V}=3.1$ and 1.7). The authors give distance moduli for all four
cases on the assumption of $H_{0}=65$. We have selected the 91 SNe\,Ia
of 2001 and later in the range of 
$3000 < \log v_{220/{\rm CMB}} < 20,000\kms$ for which we could find
the parent galaxy, position, and a sufficiently accurate redshift in
the NASA/IPAC Extragalactic Database (NED).%
% ******************************************************************
% Footnote 2
% ******************************************************************
\footnote{See http://nedwww.ipac.caltech.edu.} 
% ******************************************************************
For two SNe\,Ia (2002es and 2007qe) the available   
redshifts are discrepant; they are left out as well as the two  
deviating SNe\,Ia (2002jy and 2007bz) leaving 87 SNeIa. Their Hubble  
diagrams based on the reduction methods of \citeauthor{Guy:etal:07}
have significantly larger scatter ($\sigma_{(m-M)}=0.23$) than those
following the method of \citeauthor{Jha:etal:07}
We have adopted therefore the latter taking the mean of the moduli
from $R_{V}=3.1$ and 1.7. These means, normalized to the present
distance scale, are used to construct the Hubble diagram in
Figure~\ref{fig:01}e. 
 
     (7) \citet{Masters:etal:06} have derived relative mean cluster
distances (in $\kms$) from 21cm line width distances of 26 member
galaxies per cluster on average. Their sample of 31 clusters defines a
tight Hubble diagram with the exception of the three nearest clusters
(not shown) whose relative distances have been measured too large
\citep[see][]{TSR:08a}.  
The reason for the discrepancy, seen already in the preceding work of
\citet{Giovanelli:etal:99}, is not understood; it is apparently not
possible to apply the same selection criteria for cluster members in
nearby and distant clusters. The remaining 28 cluster distances,
normalized as before, are shown in Figure~\ref{fig:01}f.

     (8) \citet{Jorgensen:etal:96} have determined relative distances
(in $\kms$) of 10 clusters by averaging Fundamental Plane (FP)
distances of about 23 galaxies per cluster. The sample is small, but
it provides one of the few (relative) distance determinations of the
Coma cluster. The Hubble diagram with the normalized cluster distances
is in Figure~\ref{fig:01}g.
 
     The sources (1)--(8) contain 502 entries for 246 different
SNe\,Ia and 35 clusters and groups. The distance of objects with more
than one entry have been averaged. The relevant parameters of the 281
objects are listed in Table~\ref{tab:ml}.

% ********************************************************
%  Table 1: Adopted mean distances of 281 SNe Ia and clusters
% ********************************************************

     The columns that need explanation are these: 
Columns~(4) and (5) are the Galactic coordinates. Columns~(6),
(7), and (8) are the velocities relative to the Sun taken from the NED
(Col.~6), the velocities $v_{220}$ corrected for the self-consistent
Virgocentric infall model (Col.~7), and -- in case $v_{220}>3500\kms$
-- the velocities $v_{\rm CMB}$ corrected to the inertial frame of the
CMB as explained above (Col.~8). Column~(9) is the distance modulus
with the zero-point set as before by $H_{0}=62.3$. Note that the
moduli are given relative to the barycenter of the LG. 
Column~(10) is the angle in degrees from the adopted CMB apex 
$A_{\rm corr}$. Column~(11) is the difference $\Delta v_{220}$ between
the velocity $v_{220}$ and the model velocity $v_{\rm model}$
predicted by equation~(\ref{eq:carroll}) using the adopted
distance moduli from Column~(9). 
Column~(12) gives the key to the original sources. 
1: \citet{Reindl:etal:05};
2: \citet{Wang:etal:06};
3: \citet{Jha:etal:07};
4a: \citet{Wood-Vasey:etal:08};
4b: \citet{Freedman:etal:09};
5: \citet{Hicken:etal:09};
6: \citet{Masters:etal:06};
7: \citet{Jorgensen:etal:96}.

     The 281 {\em mean\/} distances in Table~\ref{tab:ml} define the
Hubble diagram in Figure~\ref{fig:01}h.
The subset of 218 objects which fall into the interval 
$3000 < v_{220/{\rm CMB}} < 20,000\kms$ scatter about the Hubble line
by $\sigma_{(m-M)}=0.15\mag$, which must be due in part to distance
errors and to peculiar velocities.

     The systematic distance difference of those SNe\,Ia occurring in
at least two of the above sources is zero by construction. Their mean
random differences are given in Table~\ref{tab:02}. The overall
average difference is $0.15\mag$ which implies a mean error of a
single distance determination of $\sim\!0.10\mag$. If this value is
subtracted in quadrature from the scatter of $0.15\mag$ observed in
Figure~\ref{fig:01}h one is left with an error, read in velocity, of
$\sigma_{\log\Delta v}=0.022$ or 5\%, which corresponds to a radial
velocity dispersion of $250\kms$ at $v_{\rm CMB}=5000\kms$ and
$430\kms$ in three dimensions. 
The value compares well with the peculiar velocity of the 
Local Supercluster of $495\kms$ toward the CMB apex
$A_{\rm corr}$ (see \S~\ref{sec:4}).

% ********************************************************
%  Table 2: Mean Random Modulus Differences Between Different Sources
% ********************************************************

     It is central to this paper to note that the free-fit 
{\em linear\/} regressions in the Hubble diagrams of
Figure~\ref{fig:01}a-h in the well occupied range of 
$3000< v_{220/{\rm CMB}}<20,000\kms$, with the slopes set out in
Table~\ref{tab:03}, are flatter than the canonical value of 0.2 
in six out of seven cases; the mean slope is flatter by $3.5\sigma$.
The reason concerns the difference between the 
world picture seen in the data and the world map as derived by the
\citet{Mattig:58} transformation between them to be described in
\S~\ref{sec:3}. 

% ********************************************************
%  Table 3: Properties of the Hubble Diagrams from Different Sources
% ********************************************************

% ******************************************************************
% 2.2. The Extension of the Hubble diagram to smaller distances
% ******************************************************************
\subsection{The Extension of the Hubble Diagram to Smaller Distances}
\label{sec:2:2}
%
% ******************************************************************
% 2.2.1. Cepheids
% ******************************************************************
\subsubsection{Cepheids}
\label{sec:2:2:1}
The uniformly reduced Cepheid distances of 30 galaxies with
$(m-M)>28.2$  were compiled in \citet{TSR:08b}.  
They are, after exclusion of the deviating NGC\,3627, plotted in
Figure~\ref{fig:02} where also the objects of
Figure~\ref{fig:01}h are repeated. 

     The Cepheids can be added directly to the Hubble diagram of
SNe\,Ia and clusters, because they all share the same zero point of
the distance scale. This is by construction, because 10 of the Cepheid
distances were used in \citet{STS:06} as the only luminosity calibrators
of SNe\,Ia. The agreement is confirmed by the statistical equality of 
$H_{0}$ from only the 29 Cepheids ($63.4\pm1.8$) by \citet{TSR:08b}
and the value derived from the SNe\,Ia ($62.3\pm1.3$) by
\citet{STS:06}.

% ********************************************************
%  Figure 2: Combined Hubble diagram
% ********************************************************

% ******************************************************************
% 2.2.2. TRGB distances
% ******************************************************************
\subsubsection{TRGB distances}
\label{sec:2:2:2}
\begin{figure}[t]
   \epsscale{0.6}
   \plotone{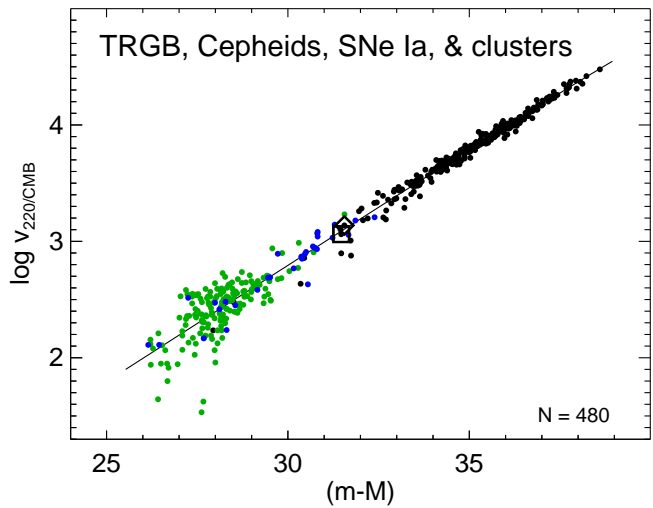}
   \caption{A combined Hubble diagram of 176 galaxies with TRGB
     distances (green) and 30 galaxies with Cepheid distances (blue);
     the data for the SNe\,Ia and clusters are repeated from
     Figure~\ref{fig:01}h. 
     For the zero-point normalization of the distance moduli
     $(m-M)$ see text. 
     The velocities are corrected for Virgocentric infall and in case
     $v_{220}>3500\kms$ for the CMB correction.
     The Virgo (square) and Fornax (diamond) clusters are plotted at
     their derived mean values $(m-M)=31.52$ and $31.65$ and 
     $v_{220}=1152$ and $1371$, respectively, from \citet{TSR:08b}.
     The fitted, slightly curved full drawn Hubble line corresponds
     to a $\Lambda$CDM model with 
     $\Omega_{\rm M}=0.3$, $\Omega_{\Lambda}= 0.7$.}
\label{fig:02}
\end{figure}

The 78, through RR~Lyr stars {\em independently\/} calibrated TRGB
distances with $(m-M)>28.2$ yield a quite local value of
$H_{0}=62.9\pm1.6$ \citep{TSR:08b}. 
The close agreement of this value with the fiducial
value of 62.3 is taken as justification to plot all 176 TRGB   
distances outside the LG into Figure~\ref{fig:02} without any
additional normalization. The combination of the SNe\,Ia and clusters
in \S~\ref{sec:2:1} with the Cepheids and TRGB distances here defines
the Hubble line in Figure~\ref{fig:02} from the lowest recession
velocities up to $30,000\kms$. 

        It could be objected that the fit of the galaxies with TRGB
distances to the more distant objects depended solely on the
similarity of the respective values of $H_{0}$, but that the agreement
was only the product of chance. Proponents of a high large-scale value
of $H_{0}>70$ are actually forced to argue in this way. To answer this
objection we have fitted separately the Hubble lines of the TRGB
objects and that of the Cepheids and SNeIa in the region of overlap, 
i.e.\ $28.2< (m-M)<31.5$. (The lower limit is chosen as elsewhere in
this paper to avoid a dominant effect of peculiar velocities). The 29
Cepheids and 19 SNe in this interval fix the intercept of the Hubble
line to within $\Delta \log v = \pm0.012$ and the 78 TRGB distances in
the same interval to within $\pm0.011$. Joining the two samples leaves
therefore an uncertainty of $\Delta \log v = \pm0.016$, corresponding
to a distance margin of $\Delta(m-M)=\pm0.08$ (or 4\% in linear
distance). Without taken regress to the actual value of $H_{0}$, the
change of $H_{0}$ between the very local value from TRGB distances and
the value on larger scales from Cepheids and SNe\,Ia can therefore be
limited to $\pm4\%$ $(1\sigma)$.

% ******************************************************************
% 2.3. Other Hubble diagrams of field galaxies and clusters
% ******************************************************************
\subsection{Other Hubble Diagrams of Field Galaxies and Clusters}
\label{sec:2:3}
Different methods to determine the distances of field galaxies have
been applied to map the velocity field. They are struck by random
distance errors of $>0.3\mag$ (except field galaxies with SNe\,Ia)
which causes severe statistical problems unless one has complete (or
fair) {\em distance-limited\/} samples. Since it is not possible to
define such samples much beyond $1000\kms$ the methods have been
applied to {\em apparent-magnitude-limited samples}. In this case the
less luminous galaxies are increasingly discriminated against as one
progresses to larger distances. The result is that the mean
luminosity of the catalogued galaxies increases with distance
(incompleteness bias). If this effect is not allowed for one derives a 
compressed distance scale with $H_{0}$ seemingly increasing with
distance. This then can lead to spurious peculiar velocities and
galaxy streamings. Of course other reasons for distance-dependent
biases exist, e.g.\ if the distance indicator depends on resolution or
if it depends on galaxy size. In all cases deviations from linear
expansion increasing systematically with distance are a sign of some
kind of bias.

     The problem of bias in the presence of large scatter is discussed
for the following four examples.

% ******************************************************************
% 2.3.1. The Hubble diagram from 21cm and optical line width distances of spirals
% ******************************************************************
\subsubsection{The Hubble diagram from 21cm and optical line width distances of spirals}
\label{sec:2:3:1}
A Hubble diagram of a complete {\em distance}-limited sample of 104
field galaxies, yet extending out to only $v_{220}=1000\kms$, has been
constructed with 21cm line width distances
\citep[][Fig.~3a]{TSR:08b}. The slope of the Hubble line is consistent
with 0.2, but the very large scatter of $\sigma_{(m-M)}=0.69\mag$ 
-- much too large to be caused by peculiar velocities -- 
prevents a rigorous test for linear expansion. Also an early
{\em magnitude}-limited sample of 217 field galaxies has been
presented by \citet{Aaronson:etal:82}. Combining 21cm line widths with
$H$ magnitudes they have derived distances which -- if cut at
$v_{220}<2500\kms$ -- define a Hubble line consistent with a slope of
0.2, but the error is substantial due to the large scatter. 

     \citet{FST:94} have analyzed a magnitude-limited sample of 1355
galaxies out to $\sim10,000\kms$ for which \citet{Mathewson:etal:92}
have collected 21cm {\em and\/} optical line widths and corrected $I$
magnitudes. The galaxies, lying in two fields towards and
perpendicular to the Great Attractor, define Hubble diagrams with
dispersions of $0.4-0.7\mag$ depending on line width, and which
clearly reveal incompleteness bias by producing slopes significantly
larger than 0.2. On the assumption of linear expansion the bias has
been corrected out by means of so-called Spaenhauer-diagrams.
The corrected distances show a local velocity anomaly in the direction
of the Great Attractor of $500\kms$ which, however, levels off
at a distance of $4000\kms$. The conclusion was that there is no
streaming extending to the Great Attractor at $v\sim4500\kms$. 

     In order to beat the large intrinsic dispersion of line width
distances and the accompanying incompleteness bias very large all-sky
samples of several thousand field and cluster galaxies have been
studied \citep{Springob:etal:07,Springob:etal:09,Theureau:etal:07}
and will be further increased in the future\citep{Masters:08}.

% ******************************************************************
% 2.3.2. Hubble diagrams from Dn-sigma and FP distances of early-type galaxies
% ******************************************************************
\subsubsection{Hubble diagrams from D$_{n}-\sigma$ and FP distances of early-type galaxies}
\label{sec:2:3:2}
\begin{figure}[t]
%   \epsscale{0.6} % <- from Fig.2 
\plotone{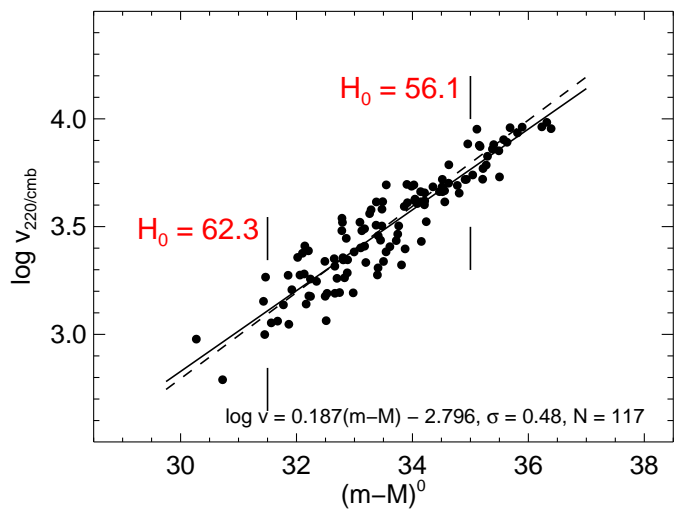}
  \caption{Hubble diagram from D$_{n}-\sigma$ distances
    \citep{Faber:etal:89}. If the value of $H_{0}$ is assumed to be
    $62.3$ at $(m-M)=31.5$ it decreases to $H_{0}=56.1$ at
    $(m-M)=35.0$. The dashed line holds for a constant value of $H_{0}$.}
  \label{fig:DnSigma}
\end{figure}

\citet{Faber:etal:89} have derived D$_{n}-\sigma$ distances (in $\kms$)
of 317 E and S0 galaxies out to $\sim\!10,000\kms$.
The corresponding Hubble diagram  has a scatter of
$\sigma_{(m-M)}=0.69\mag$, which is reduced to
$\sigma_{(m-M)}=0.50\mag$ by correcting for what the authors call
``Malmquist correction'', but which is actually a correction for the
population size of the aggregate from which a galaxy is drawn. 
In their Table~4 they list {\em mean\/} distances of 59 clusters and
groups and distances of 58 single galaxies which define the Hubble
diagram shown in Figure~\ref{fig:DnSigma}. The scatter is still large
($0.48\mag$) and the slope ($0.187\pm0.07$) is flatter than the
expected value of $0.2$. This implies that if $H_{0}$ is assumed to be
$62.3$ at $(m-M)=31.5$ it decreases to $H_{0}=56.1$ at $(m-M)=35.0$.
The decrease of $H_{0}$ is contrary to what is expected from an
incompleteness bias; it may be that the specific ``Malmquist
correction'' of the authors leads to an overcorrection of the
distances.

% ********************************************************
% Figure 3: Hubble diagram - Dn-sigma
% ********************************************************

     All of \citet*{Faber:etal:89} galaxy distances were analyzed in a
number of publications some of which preceded the paper by
\citeauthor{Faber:etal:89} 
\citep[e.g.][]{Dressler:87,Faber:Burstein:88,Lynden-Bell:etal:88,Faber:etal:88,Burstein:90}.
The authors, taking the distances at face value, concluded that a
coherent large-scale influx existed into a group of galaxy clusters
centered on Abell~3627, called the ``Great Attractor'', at a distance
of $\sim\!4500\kms$, whose dominant attraction would be the main cause
for the Local Group's motion with respect to the CMB. 
-- First doubts about the r{\^o}le of the
``Centaurus Concentration'' (synonymous for Great Attractor) came from
\citet{Lynden-Bell:etal:89}, who concluded from the optical dipole
from galaxy catalogs that most of the Local Group's motion is caused
from within $3500\kms$ and that the more distant cluster concentration
about Abell 3627 was only a minor contributor. Additional evidence for
this view is given in \S~\ref{sec:5}.

     Mean Fundamental Plane (FP) distances of 85 clusters with 
$v_{\rm CMB}=5000-20,000\kms$ have also been derived by
\citet{Colless:etal:01}, but they are based on average on only
$\sim\!5$ cluster members. The resulting Hubble diagram has a large
scatter of $\sigma_{(m-M)}=0.34\mag$ and the Hubble line has too flat
a slope ($0.165$). For these reasons the data have not been used for the
linearity test. 

     Forthcoming bias-corrected FP distances of up to $\ga10,000$
galaxies in the Southern sky are announced
\citep{Smith:etal:04,Jones:etal:04,Springob:etal:10}; they shall serve
to map the velocity field of the local universe out to $\la20,000\kms$.

% ******************************************************************
% 2.3.3. The Hubble diagram from surface brightness fluctuation (SBF) distances
% ******************************************************************
\subsubsection{The Hubble diagram from surface brightness fluctuation (SBF) distances} 
\label{sec:2:3:3}
\begin{figure}[t]
%   \epsscale{0.6} % <- from Fig.2 
\plotone{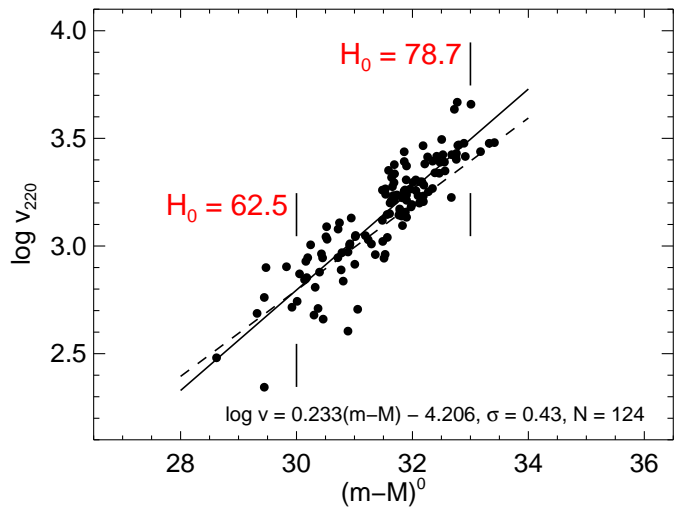}
  \caption{Hubble diagram from SBF distances
    \citep{Tonry:etal:01}. The published distances give $H_{0}=62.5$
    at $(m-M)=30.0$ and $H_{0}=78.7$ at $(m-M)=33.0$.
    The dashed line holds for a constant value of $H_{0}$.} 
  \label{fig:SBF}
\end{figure}

SBF distances of 124 mainly E and S0 galaxies have been published by  
\citet{Tonry:etal:01}. Their Hubble diagram is shown in
Figure~\ref{fig:SBF}. The scatter of $\sigma_{(m-M)}=0.43\mag$ 
is large; even for this relatively nearby sample 
($v_{220}(\mbox{median}) = 1626\kms$) only part of the scatter can be
attributed to peculiar motions. Yet the most striking is the steepness
of the Hubble line with slope 0.233; it implies values of $H_{0}=62.5$
at $(m-M)=30.00$ and of $H_{0}=78.7$ at $(m-M)=33.00$. The seeming
non-linearity of the expansion field is impossible and must be due to
some bias of the data, which may be caused by a selection effect
depending on apparent-magnitude (incompleteness bias) or some other
distance-dependent bias of this sensitive method (perhaps due to
incomplete removal of non-stellar images). 
The consequence is that the method should not be used to test the
character of the expansion field nor for the determination of reliable
galaxy distances. 

% ********************************************************
% Figure 4: Hubble diagram - SBF
% ********************************************************

% ******************************************************************
% 2.3.4. The Hubble diagram from planetary nebula (PNLF) distances
% ******************************************************************
\subsubsection{The Hubble diagram of planetary nebula (PNLF) distances} 
\label{sec:2:3:4}
\begin{figure}[t]
%   \epsscale{0.6} % <- from Fig.2 
\plotone{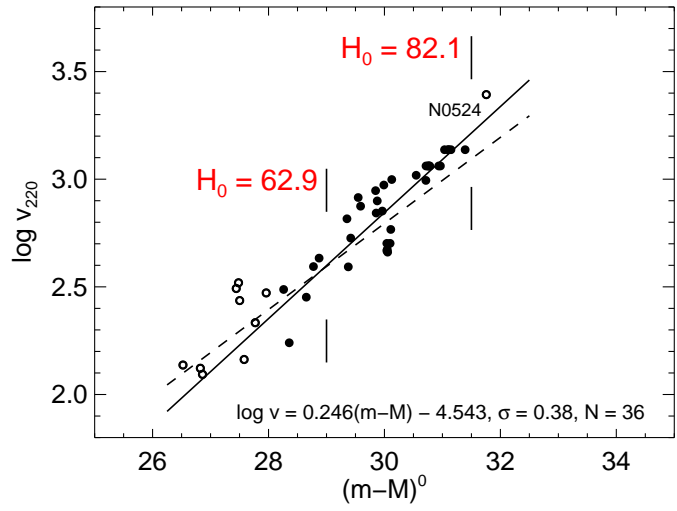}
  \caption{Hubble diagram from PNLF distances. The published distances
    give $H_{0}=62.9$ at $(m-M)=29.0$ and $H_{0}=82.1$ at
    $(m-M)=31.5$. The dashed line holds for a constant value of
    $H_{0}$. The open symbols are not used for the fit.}
  \label{fig:PNLF}
\end{figure}

Galaxy distances derived from the bright tail of the luminosity
function of the shells of planetary nebulae in the light of the
$\lambda\,5007\,${\AA} line have been compiled from
\citet{Ciardullo:etal:02}, \citet{Feldmeier:etal:07},
\citet{Herrmann:etal:08}, and the NED.
Of the resulting galaxies 46 lie outside the LG. They define a Hubble
diagram as shown in Figure~\ref{fig:PNLF}. A free fit through the 36
points beyond $(m-M)=28.2$ (omitting NGC\,524 with only a lower limit
to its distance) leads to a Hubble line with large scatter
($\sigma_{(m-M)}=0.43\mag$) and very steep slope. The non-linearity is
so pronounced that $H_{0}$ increases from 62.9 to 82.1 as one goes
from $(m-M)=29.0$ to 31.5! The reason may be that the brightness of
the brightest planetary nebulae depends on galaxy size
\citep{Bottinelli:etal:91,Tammann:93}. Whatever the reason, the
luminosity function of planetary nebulae is not a useful distance
indicator. 

% ********************************************************
% Figure 5: Hubble diagram - PNLF
% ********************************************************

% ******************************************************************
% 3. The test for linearity of the local expansion field
% ******************************************************************
\section{\MakeUppercase{The Test for Linearity of the Local Expansion Field}}
\label{sec:3}
As said at the end of \S~\ref{sec:1:2}, the test for linearity -- the
test if $H_{0}$ is or is not a function of distance in the world map
-- is model-dependent because it depends on transforming the world
picture into the world map. This can only be done by using the $R(t)$
scale factor which depends on the model, and by taking account of the
streaming motions where they exist.

     The problem and its evident degeneracy is illustrated by the
deviation of the slopes in Table~\ref{tab:03} from $0.200$ which is
the value if the world {\em map\/} describes a homogeneous and
isotropic universe in the mean. The naive interpretation of the mean
slope of $0.193\pm0.002$ would be that the Hubble constant 
{\em decreases\/} outward such that if $H_{0}=62.3$ locally at say 
$v=1000\kms$, then the value at $10,000\kms$ would be 58.0.  

     However, such a conclusion is incorrect. It ignores the
transformation of the world picture into the world map via a
Mattig-like equation where a second parameter is required in the
theoretical, model dependent equation for the Hubble diagram.        

     We test first in \S~\ref{sec:3:1} for linearity in
case of a fixed value of $q_{0}$ and then investigate in
\S~\ref{sec:3:2} the range of $q_{0}$ for which linearity cannot be
excluded using only the local data with $z<0.1$.

% ******************************************************************
% 3.1 The test for linearity using a world model with O_M=0.3, O_L=0.7
% ******************************************************************
\subsection{\boldmath The Test for Linearity Using a World Model with 
    $\Omega_{\rm M}=0.3, \Omega_{\Lambda}=0.7$}
\label{sec:3:1}
In a first step of the linearity test we adopt the values of
the ``concordance model'' of $\Omega_{\rm M}=0.3$,
$\Omega_{\Lambda}=0.7$, which corresponds to $q_{0}=-0.55$.

The velocity residuals $\Delta\log v$ of the objects in
Figure~\ref{fig:02} from the Hubble line calculated with
equation~(\ref{eq:carroll}) are plotted against
the adopted distance moduli in Figure~\ref{fig:06}a. 
The scatter of the residuals increases towards smaller distances due
to the relatively larger contribution of peculiar motions to the
recession velocities. The distribution of the points is as flat as can
be expected, giving a slope of $0.000\pm0.001$. 
This implies that the value of $H_{0}$ remains constant to within
$\pm2.3\%$ over a modulus interval of $\Delta(m-M)=10$, corresponding
to a distance factor of 100.  

% ********************************************************
%  Figure 6: Deviations Delta log v from the adopted Hubble line
% ********************************************************

     In order to test for {\em local\/} distance-dependent variations
of $H_{0}$ the data of Figure~\ref{fig:06}a are averaged per bins of
width  $\Delta(m-M)=0.5\mag$ and shown in Figure~\ref{fig:06}b. 
The 21 averaged residuals between $(m-M)=28.0$ and 38.0 are about
normally distributed and have an average rms deviation of 
$\sigma_{\Delta\log v}=0.012$ from the adopted horizontal line, which
restricts the local variations of $H_{0}$ to $2.8\%$ on average.
Individual intervals with 10 or less objects deviate by up to 7\%.
Shifting the boundaries of the bins does not change this result. 

     Earlier suggestions of $H_{0}$ to decrease gradually  by 5\% out
to $18,000\kms$ \citep{Tammann:99} cannot be maintained in the light
of the present, much increased data, which also deny the so-called
``Hubble Bubble''. This feature was proposed as a drop of $H_{0}$ by 
$6.5\pm2.0\%$ beyond a distance of $\sim\!7200\kms$ (corresponding to
$(m-M)=35.3$ in the adopted distance scale) by \citet{Zehavi:etal:98}
and \citet{Jha:etal:07}. This drop has been taken as support for a
local overdensity and for a dark-energy-free cosmology
\citep{Wiltshire:08}. Yet \citet{Giovanelli:etal:99},
\citet{Conley:etal:07}, and  \citet{Hicken:etal:09} have questioned
the result as we do here.
Indeed comparing the bin in Figure~\ref{fig:06}b just preceding the
break with the one following the break -- either bin containing more
than 30 objects -- shows the more distant bin to lie {\em higher\/}
by $\delta(\Delta\log v)=0.005\pm0.008$ or $H_{0}$ to be 
{\em larger\/} by $1.3\pm1.7\%$. 
This rejects the Hubble Bubble at the level of 4 sigma.
-- It is also noted that the five intervals within $32.5<(m-M)<35.0$
and the five intervals within $35.5<(m-M)<38.0$,
embracing the break point and comprising a total of more than 100
objects each, give an increase of $H_{0}$ with distance of
$0.5\pm1.2\%$ instead of the putative decrease of $\sim\!6.5\%$. This
rejects a persistent decrease of $H_{0}$ at an even higher level of
significance.

% ******************************************************************
% 3.2 The model dependence of the linearity test
% ******************************************************************
\subsection{The Model Dependence of the Linearity Test}
\label{sec:3:2}
It is interesting to ask for the sensitivity of the test for linearity
for a range of world models.

% ******************************************************************
% 3.2.1 The linearity test for different values of O_M,O_L
% ******************************************************************
\subsubsection{The linearity test for different values of
  $\Omega_{\rm M}, \Omega_{\Lambda}$}
\label{sec:3:2:1}
We first consider two {\em flat\/} $\Lambda$CDM models with rather
extreme values of ($\Omega_{\rm M}, \Omega_{\Lambda}$) = (0.6, 0.4)
and (0.1, 0.9) corresponding to $q_{0}=-0.10$ and $-0.85$,
respectively. The differences between the moduli $(m\!-\!M)_{\rm
  model}$ for the two trial models from equation~(\ref{eq:carroll})
and the moduli for the standard model under \S~\ref{sec:3:1} are
listed in Table~\ref{tab:q0}, Columns (3) \& (4) for various values of
$z$.  The differences are shown as smooth curves in
Figure~\ref{fig:07}, where the differences $\Delta(m\!-\!M)$ are
plotted against $\log z$. Also shown are the mean values of 
$\Delta(m\!-\!M)=(m\!-\!M)_{\rm obs}-(m\!-\!M)_{\rm standard\,model}$ 
of all objects in Figure~\ref{fig:06} averaged over intervals of 
$\Delta\log z=0.3$ and plotted in steps of $\Delta\log z=0.1$. The
conclusion is that the averaged data points at larger distances fit
reasonably within the boundaries of the two trial models. This means 
that linearity holds for all flat models with parameters between the
two trial models. The weak restriction on the parameter space is no
surprise in view of the short leverage in $z$ of the data used.

% ********************************************************
%  Table 4: Delta(m-M) for different q_0
% ********************************************************

% ********************************************************
%  Figure 7: Modulus differences Delta(m-M) vs. log z
% ********************************************************

% ******************************************************************
% 3.2.2 The linearity test for different values of q_0
% ******************************************************************
\subsubsection{The linearity test for different values of $q_{0}$}
\label{sec:3:2:2} 
We now relax the condition of a flat model and specify the trial
models only by $q_{0}$. For this purpose we recall the derivation of
the theoretical Hubble diagram of $\log z$ vs. apparent magnitude
using a standard candle in various homogeneous, isotropic universes.   

     \citet*{Robertson:38} equation that relates the apparent
luminosity, $l$, with the metric distance at the present instant of
cosmic time is
\begin{equation}
           l = L[(4\pi R_{0}r)^{-2} (1+z)^{-2}]             % (2)
\label{eq:01}
\end{equation}
where $L$ is the absolute luminosity and $R_{0}r$ is the metric
distance where $r$ is the constant (for all time) dimensionless metric
distance, and $R_{0}$ is given by the Mattig solution (for
$\Lambda=0$) as  
\begin{equation}
R_{0}r = \frac{c}{H_{0}q_{0}^{2}(1+z)}
         [zq_{0}+(q_{0}-1)\{-1 + (2q_{0}z+1)^{1/2}\}].      % (3)
\label{eq:02}
\end{equation}
 
     Combining equations~(\ref{eq:01}) and (\ref{eq:02}) gives the
equation for the theoretical Hubble diagram to be
\begin{equation}
    (m\!-\!M) = 5\log q_{0}^{-2}[zq_{0} + 
            (q_{0}^{-1})\{-1 +(2q_{0} +1)^{1/2}\}] + C,     % (4)   
\label{eq:03}
\end{equation}
where $C$ contains the Hubble constant and the intrinsic mean absolute
magnitude of the distance indicator.  

     The series expansion of equation~(\ref{eq:03}), is
\begin{equation}
       (m\!-\!M) = 5\log z + 1.086(1-q_{0})z
       +1/6(2-q_{0}-3q_{0}^{2})z^{2} + O(z^{3}) + C.        % (5)
\label{eq:04}
\end{equation}
For small values of $z$ equation~(\ref{eq:04}) is also valid if
$\Lambda\ne0$.  The effect of the third term is $\le0.003$ for 
$q_{0}>-1$ and $z<0.1$.  

     The values of $(m\!-\!M)$ are calculated from
equation~(\ref{eq:04}) for four trial values of 
$q_{0}=0.00, -0.20, -0.80$, and $-1.00$ and for various values of $z$.
Subtracting from these values the modulus $(m\!-\!M)$ resulting from
$q_{0}=-0.55$ and the corresponding values of $z$, one obtains the
values $\Delta(m\!-\!M)_{\rm model}$ listed in Table~\ref{tab:q0}
(Col. 5-8) and shown in Figure~\ref{fig:07} as smooth dashed lines. 
As in \S~\ref{sec:3:2:1} the more distant data points lie within the
boundaries for $q_{0}=0.00$ and $-1.00$. The conclusion is that
$q_{0}$ is probably negative irrespective of the value of $\Lambda$.

% ******************************************************************
% 3.3 The linearity of the expansion at large scales 
% ******************************************************************
\subsection{The Linearity of the Expansion at Large Scales} 
\label{sec:3:3}
If linear expansion -- albeit for a considerable range of parameters
-- was found in the foregoing from a sample arbitrarily cut at $z<0.1$
($30,000\kms$), there is sufficient overlap with other SN\,Ia samples
extending to higher redshifts to carry the linearity test to $z>1$
(\S~\ref{sec:1:3}). 

     The main conclusion of \S~\ref{sec:3} is that the cosmic
expansion is linear on all scales. This holds surprisingly, if
corrections are applied for the local Virgocentric flow and for the
bulk motion of the Local Supercluster toward the warm pole of the CMB 
(see \S~\ref{sec:4}), down to distances of $300\kms$ ($\sim5\;$Mpc).

% ******************************************************************
% 4. The size of the volume co-moving towards the CMB apex
% ******************************************************************
\section{\MakeUppercase{The size of the volume co-moving towards the CMB apex}}
\label{sec:4}
As evidenced by the dipole of the CMB radiation the LG partakes
of a bulk motion in addition to the Virgocentric vector. The
dipole, first received with disbelief, was discovered by
\citet{Henry:71}, \citet{Corey:Wilkinson:76}, and
\citet{Smoot:etal:77}.  
After attempts to explain the dipole as a primordial effect
\citep{Gunn:88,Paczynski:Piran:90} it soon became clear that it could
be caused only by a local motion which, seen from the Sun, amounts to 
$369\pm2\kms$ towards the observed apex $A_{\rm obs}$ at 
$l=263.86\pm0.04$, $b=48.24\pm0.10$ 
\citep{Bennett:etal:96,Hinshaw:etal:07} or, if translated to the
barycenter of the LG 
following \citet{Yahil:etal:77}, to  $v_{\rm LG}=626\pm30\kms$ 
towards $l=276\pm2$, $b=30\pm2$.
Subtracting the Virgocentric infall vector from $v_{\rm LG}$ one
obtains a velocity of the LG of $v_{\rm CMB}=495\pm25\kms$ 
towards the corrected CMB apex $A_{\rm corr}$ at $l=275\pm2$, $b=12\pm4$
\citep[cf.\ also][]{Sandage:Tammann:84}. It was clear from the
beginning that a velocity of such size must comprise a very large
volume including the Virgo cluster and extending to at least
$1500\kms$ because otherwise two galaxies at roughly equal distances
in the apex and antapex direction would differ in redshift by
$990\kms$! The conclusion is that the whole Local Supercluster moves
more or less coherently towards the CMB apex. 

% ********************************************************
% Figure 8: Hypothetical Hubble diagram - Apex vs. Antapex
% ********************************************************

     Figure~\ref{fig:hypoHD} shows the aspect of the Hubble
diagram if the Galaxy were the only object moving towards the CMB pole
with a velocity of $495\kms$. In that case all other nearby galaxies,
being at rest in co-moving coordinates with respect to the CMB, would
exhibit large peculiar motions as seen from the LG and fill the
Hubble diagram within the wide curved envelopes as shown in the
Figure. The assumption, besides being unphysical, is clearly
contradicted by the concentration of the observed galaxies 
(repeated from Fig.~\ref{fig:02}) towards the center line of the
diagram. Most galaxies within say $1000\kms$ must share the CMB
motion of the Galaxy. Beyond this point the diagram loses its
diagnostic power because the vector of $495\kms$ is drowned in the
recession velocities of more distant galaxies and in the natural
scatter.    

     The convergence radius of the Local Supercluster, i.e.\ the
distance where field galaxies merge into the inertial frame of the CMB
has been determined by several authors, some of which are compiled in
Table~\ref{tab:LSconvergence}. 
The size and direction of the bulk motion from the different
authors are not listed in Table~\ref{tab:LSconvergence}, because some
authors postulate additional velocity vectors on larger scales to
explain the full CMB dipole motion. One of the reasons -- besides
systematic distance errors which always lead to large-scale motions --
may be that they have not allowed for Virgocentric infall vectors and
that they have compared their results with the CMB apex $A_{\rm obs}$
instead of $A_{\rm corr}$ (i.e. after correction of the
local Virgocentric infall vector of $220\kms$).

% ********************************************************
%  Table 5: Convergence Depth of the Local Supercluster
% ********************************************************

     A direct determination of the size of the Local Supercluster and
its bulk motion is obtained when one plots the velocity residuals from
the Hubble line $\Delta v_{220}= v_{220} - v_{\rm model}$ against the
angle $\alpha$ from the apex $A_{\rm corr}$. 
Galaxies from Figure~\ref{fig:02} are divided into two bins, 
$500<v_{220}<3500$ and $3500<v_{220}<7000\kms$ and their apex diagrams
are shown in Figure~\ref{fig:apex}a,b. The result is striking.
The distribution of the nearer sample is flat and the more distant
sample reflects a local peculiar velocity of $448\pm73\kms$ 
in satisfactory agreement with the corrected CMB dipole motion of
$495\pm25\kms$. Clearly the majority of galaxies within
$\sim\!3500\kms$ share the coherent bulk motion of the Local
Supercluster, whereas the galaxies beyond this limit are in rest with
respect to the inertial frame of the CMB. 

% ********************************************************
%  Figure 9: Velocity residuals
% ********************************************************

     The scatter in Figure~\ref{fig:apex}b is significantly larger
than in Figure~\ref{fig:apex}a. This is because the distance errors of
$\ga6\%$ predict velocities with errors increasing with distances
such as to overwhelm eventually the signal from streaming
velocities. For this reason the apex diagrams should not be carried
much beyond $7000\kms$. 

Aiming for a still better determination of the convergence length, the
objects of Figure~\ref{fig:apex} in the range $2400<v_{220}<4800\kms$
have been re-binned in $1000\kms$ intervals and are plotted in steps
of $200\kms$ in Figure~\ref{fig:apex2}. The nearer samples in panel a)
and b) show already a marginal slope suggesting that the Local
Supercluster begins to peter out at even smaller velocities
(distances) than $3500\kms$. Panel c) centered at $3300\kms$ support
that conclusion at the $1.6\sigma$ level, while panel d) centered at
$3500\kms$ shows a highly significant apex motion. Finally the more
distant panels e)$-$h) are fully consistent with the expected asymptotic
apex motion of $495\pm25\kms$.

% ********************************************************
% Figure 10: Velocity residuals (4x2)
% ********************************************************

     The conclusion is that the convergence distance is reached 
close to $3500\kms$. 
The exact convergence distance depends presumably on direction, but for
a corresponding test the large, yet still restricted sample is not
adequate. 

    For visualization the 90 objects in Figure~\ref{fig:apex}b (i.e.\
with $3500<v_{\rm 220}<7000\kms$) are plotted in an Aitoff projection
(Fig.~\ref{fig:aitoff}). The association of the objects with positive
and negative velocity residuals with their respective apices is
striking. The eccentric position of the apices within their associated
objects is due to the paucity of objects in the zone of avoidance.
The lopsided distribution of the available objects about the apices is
also the reason why we have not independently solved for the direction
of the apex, but have taken the CMB dipole direction as given.

% ********************************************************
%  Figure 11: Aitoff Diagram
% ********************************************************

% ********************************************************
%  Figure 12: Flow diagram
% ********************************************************

     The flow pattern of the 144 sample objects with
$500<v_{220}<10,000\kms$ and $|\beta|<45^{\circ}$ is shown in
Figure~\ref{fig:flow}. 
Their distances and velocity residuals $\Delta v_{220}=v_{220}-v_{\rm
  model}$ are projected on the ``apex plane'' which is tilted against
the Galactic plane by $12^{\circ}$ and rotated such that the CMB apex
$A_{\rm corr}$ has the new coordinates $\lambda=270^{\circ}$,
$\beta=0^{\circ}$. 
The objects within $3500\kms$ have small
velocity residuals, whereas the more distant objects have larger
velocity residuals on average.
The distribution of the available objects beyond $3500\kms$, although
not forming a complete nor objectively selected sample, is far from
random. Of the 18 objects in the $45^{\circ}$ sector about the apex
direction 12 have negative and 6 have positive velocity residuals. The
chance probability for this distribution is $12\%$. Of the 29 objects
in the corresponding sector about the antapex direction 25 have
positive and 4 have negative velocity residuals. In this case the
chance probability is only $P=0.01\%$. The bulk motion of the Local
Supercluster is hence reflected in Figure~\ref{fig:flow} at a high
level of significance. Consistent with this is the nearly random
distribution of the velocity residuals in the two perpendicular
sectors about $\lambda=0^{\circ}$ and $\lambda=180^{\circ}$.

% ******************************************************************
% 5. Results and conclusions
% ******************************************************************
\section{\MakeUppercase{Results and Conclusions}}
\label{sec:5}
Exceptionally accurate relative distances are compiled from eight
sources in \S~\ref{sec:2:1}. They are based on SN\,Ia magnitudes and
on mean 21cm line widths and the Fundamental Plane in case of
clusters. They are normalized to an arbitrary value of $H_{0}$ of 62.3
without loss of generality. 
Accurate TRGB and Cepheid distances were added in
\S~\ref{sec:2:2}. The final Hubble diagram with 480 objects extends
from 300 to $30,000\kms$ and has a scatter of 
only $0.15\mag$ (8\% in linear distance) beyond $3000\kms$; the
increase of the scatter at shorter distances, due to peculiar motions,
is compensated by large-number statistics. 

     The expansion of space is as linear as can be measured after
allowance is made for a Virgocentric flow model and for the CMB motion
of the Local Supercluster. Over the range of 300 to $30,000\kms$ the
value of $H_{0}$ does not change systematically by more than
$\pm2.3\%$. At poorly populated distances $H_{0}$ may locally vary by
7\%. But the so-called Hubble Bubble suggesting a sudden decrease of
$H_{0}$ by 6.5\% -- be it a local dip or a persistent feature beyond
-- at the particularly well occupied distance around $7200\kms$ is
excluded by $4\sigma$ or more.

\begin{figure}[t] % [p]
  \epsscale{0.60}
   \plotone{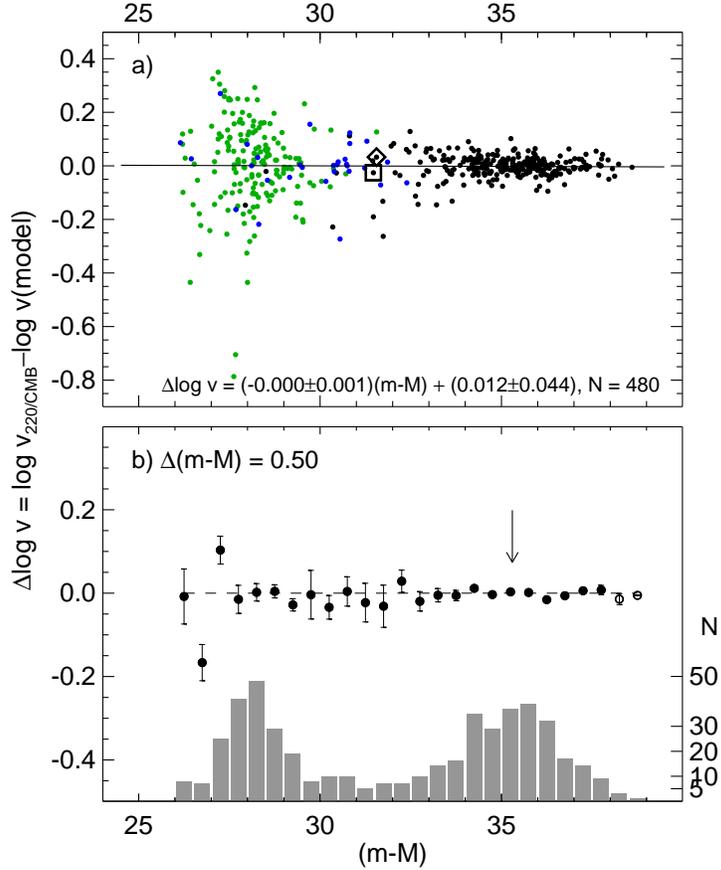}
    \vspace*{0.4cm}
   \caption{Panel a): The deviations $\Delta\log v$ from the adopted
   Hubble line of the objects in Figure~\ref{fig:02} are plotted
   versus distance modulus. 
   The full drawn line is a regression to all points. Note its
   flatness, allowing the average value of $H_{0}$ to vary
   systematically as a function of distance by not more than
   2.3\% ($1\sigma$) over a distance interval of a factor of 100.
   Symbols as in Figure~\ref{fig:02}.
   Panel b): same as a), but the distances of objects within $0.50\mag$
   bins have been averaged. Any deviations from the adopted Hubble
   line are hardly significant. Error bars are only shown if they are
   larger than the symbols. Bins with less than 5 elements are shown
   as open symbols. 
   The arrow at $(m-M)=35.3$ (corresponding to $v\sim\!7200\kms$) is
   where \citet{Zehavi:etal:98} and \citet{Jha:etal:07} have suggested
   a drop of $H_{0}$ by 6.5\% (the so-called ``Hubble Bubble'') which
   is denied by the present data. The histogram at the bottom of the
   diagram shows the number of objects per bin.}
\label{fig:06}
\end{figure}

\begin{figure}[t]
   \epsscale{0.6}
   \plotone{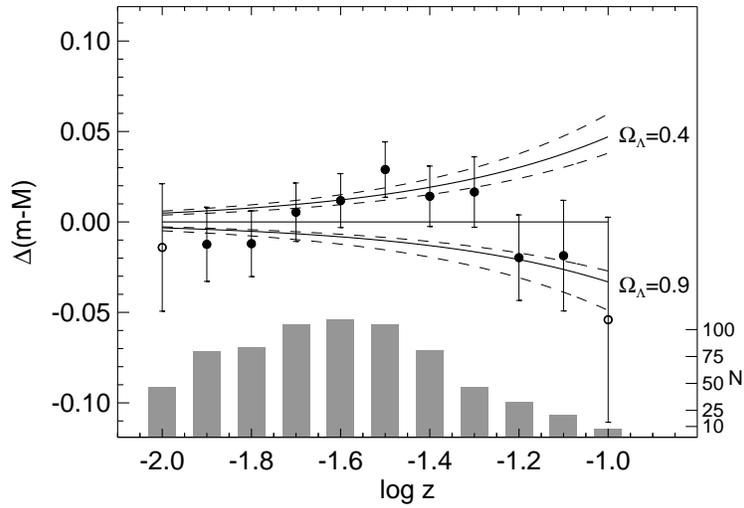}
   \caption{The modulus differences $\Delta(m-M)$ vs. the
   model-independent value of $\log z$ for different trial models as
   compared to the standard model specified by  ($\Omega_{\rm M},
   \Omega_{\Lambda}$) = (0.3, 0.7) or $q_{0}=-0.55$. The full curved
   lines correspond to two flat models with ($\Omega_{\rm M},
   \Omega_{\Lambda}$) = (0.6, 0.4) and (0.1, 0.9), respectively. 
   The dashed lines stand for four models specified by only
   $q_{0}=0.00$, $-0.20$, $-0.80$, and $-1.00$ (top to bottom).
   Running means of the data from Figure~\ref{fig:06} are shown for
   intervals of $\log z=0.3$ in steps of $\Delta\log z=0.1$. 
   The histogram shows the number of objects per $\log z$ interval of
   0.1.}   
\label{fig:07}
\end{figure}

\begin{figure}[t]
   \epsscale{0.6}
\plotone{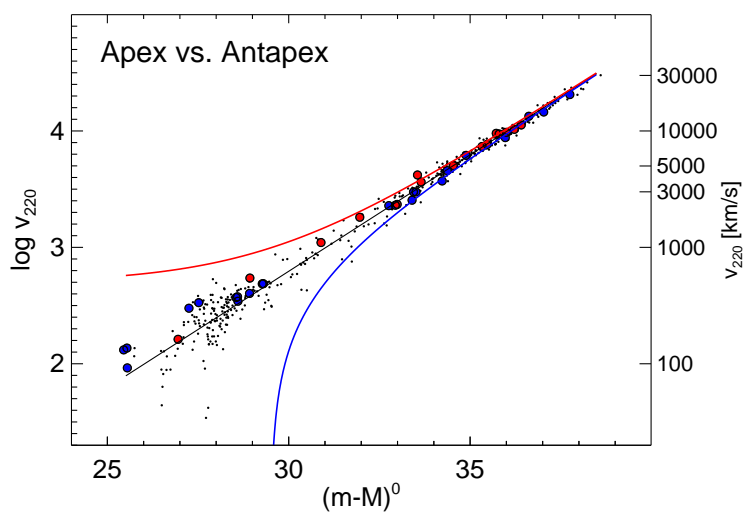}
  \caption{A hypothetical Hubble diagram in case that neighboring
    galaxies were in rest with the CMB radiation and 
    had no streaming motion towards the apex of the CMB. In that case
    the galaxies would occupy the space between the two curved
    envelopes. The observed galaxies within $\sim1000\kms$ center
    strongly towards the center line of the diagram, have hence small
    peculiar velocities and must share the Galactic CMB motion.
    The simple test fails beyond $\sim1000\kms$. Small dots are the
    objects of Figure~\ref{fig:02}. Red and blue points lie within 30
    degrees from the apex and antapex, respectively; in the
    hypothetical case they would lie near to the envelope lines.}   
  \label{fig:hypoHD}
\end{figure}

\begin{figure}[t]
   \epsscale{0.6}
\plotone{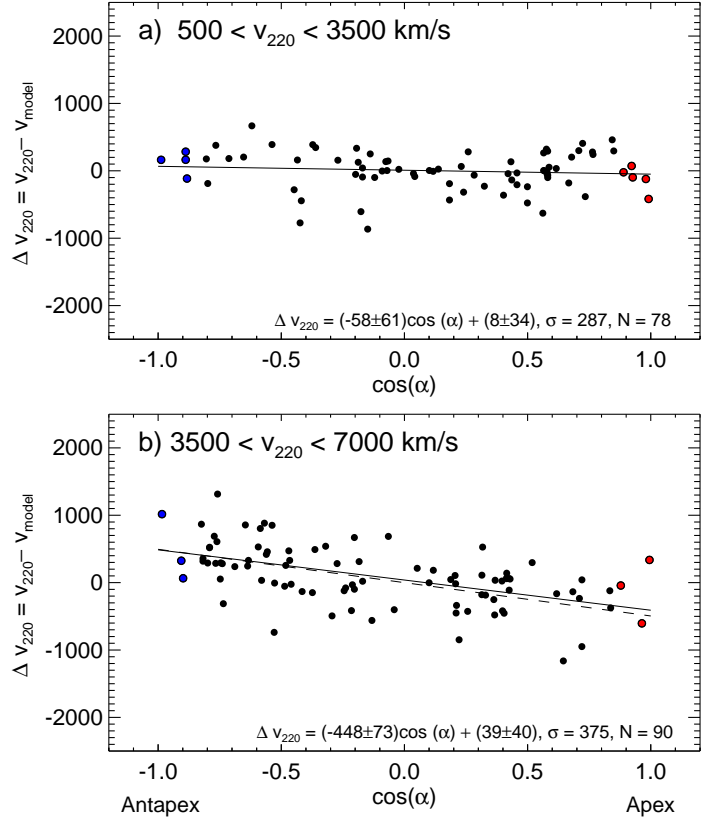}
\caption{Velocity residuals $\Delta\log v_{220}$ from
  Table~\ref{tab:ml} in function of $\cos(\alpha)$, where $\alpha$ is
   the angle away from the CMB apex $A_{\rm corr}$, for 
   a) $500 < v_{220} < 3500$   and  
   b) $3500 < v_{220} < 7000\kms$. 
   Galaxies with $v_{220} < 3500$ move coherently within the Local
   Supercluster. Galaxies with $v_{220} > 3500$ reflect a systematic
   velocity of the Local Supercluster of $448\pm73\kms$ toward the CMB
   apex $A_{\rm corr}$. The velocity agrees with $495\pm25\kms$ as
   derived from the CMB data.
   Objects within $30^{\circ}$ from the apex (antapex) are shown in
   red (blue). The dashed line in panel b) holds for objects in rest
   relative to the CMB. Symbols as in Figure~\ref{fig:hypoHD}.}
  \label{fig:apex}
\end{figure}

\begin{figure}[t] % [p]
\epsscale{0.92}
\plotone{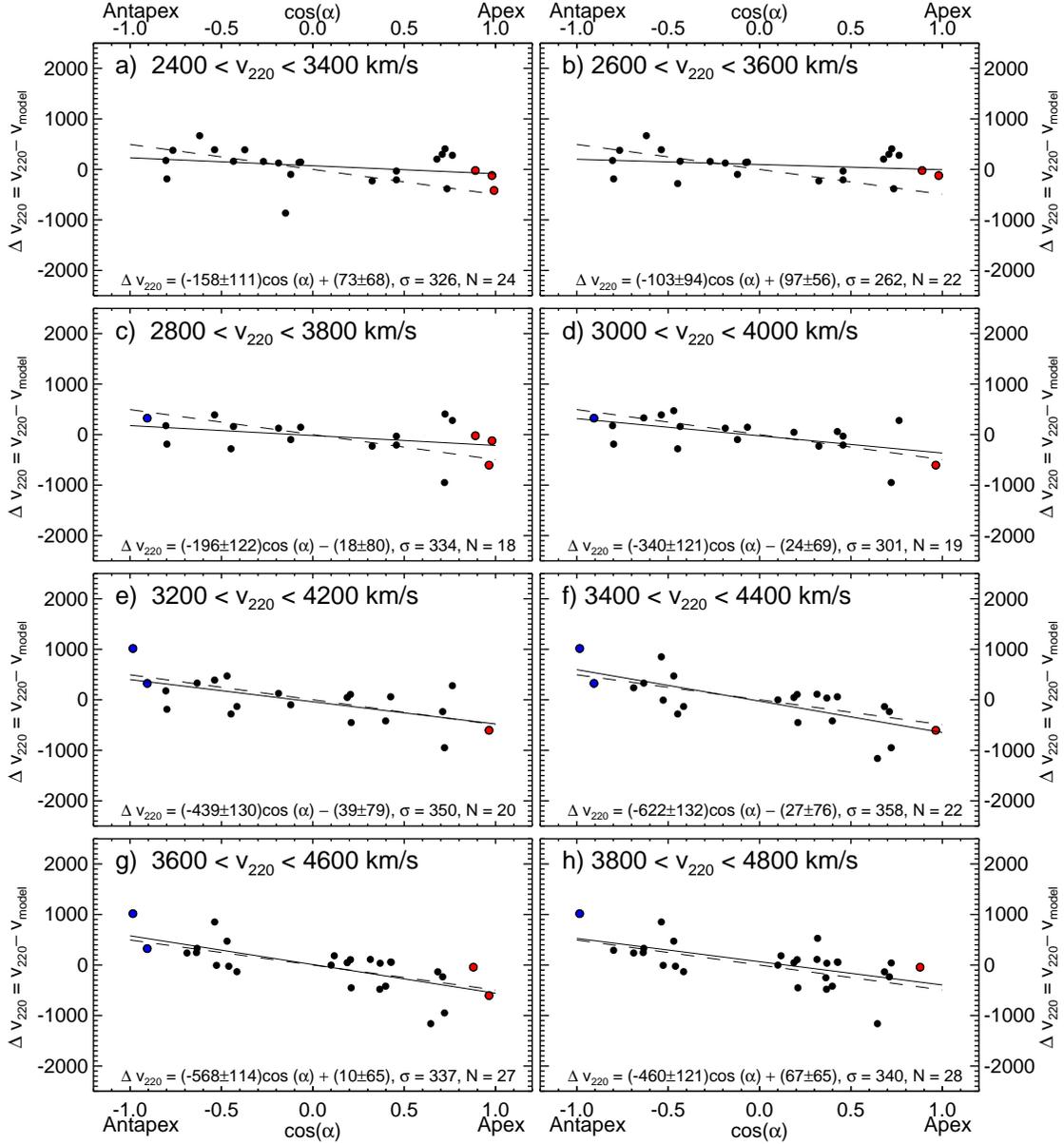}
  \caption{Same as Figure~\ref{fig:apex}, but for velocity intervals
    with a width of only $1000\kms$ and plotted every $200\kms$. The
    expected asymptotic apex velocity of $495\kms$ is reached in panel
    d) or e), i.e.\ at $v_{220}=3500\kms$.}  
  \label{fig:apex2}
\end{figure}

\begin{figure}[t] % [p]
   \epsscale{0.92}
\plotone{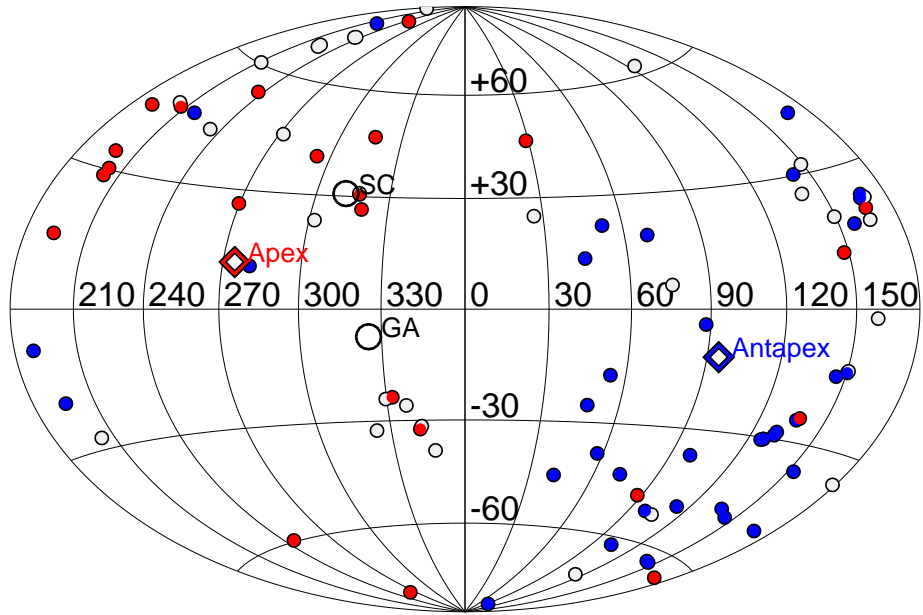}
\caption{An Aitoff projection in Galactic coordinates of the objects 
  with $3500<v_{\rm CMB}<7000$. Objects with residual velocities
  $\Delta v<150\kms$ are shown in red, those with $\Delta v>150\kms$
  in blue. Objects with residuals $-150<\Delta v<150\kms$ are shown as
  open symbols. The positions of the corrected apex and antapex (after
  subtraction of the Virgocentric infall vector of the LG)
  as well as of the Great Attractor (GA) and the Shapley
  Concentration (SC) are indicated.}
  \label{fig:aitoff}
\end{figure}

\begin{figure}[t] % [p]
   \epsscale{0.92}
\plotone{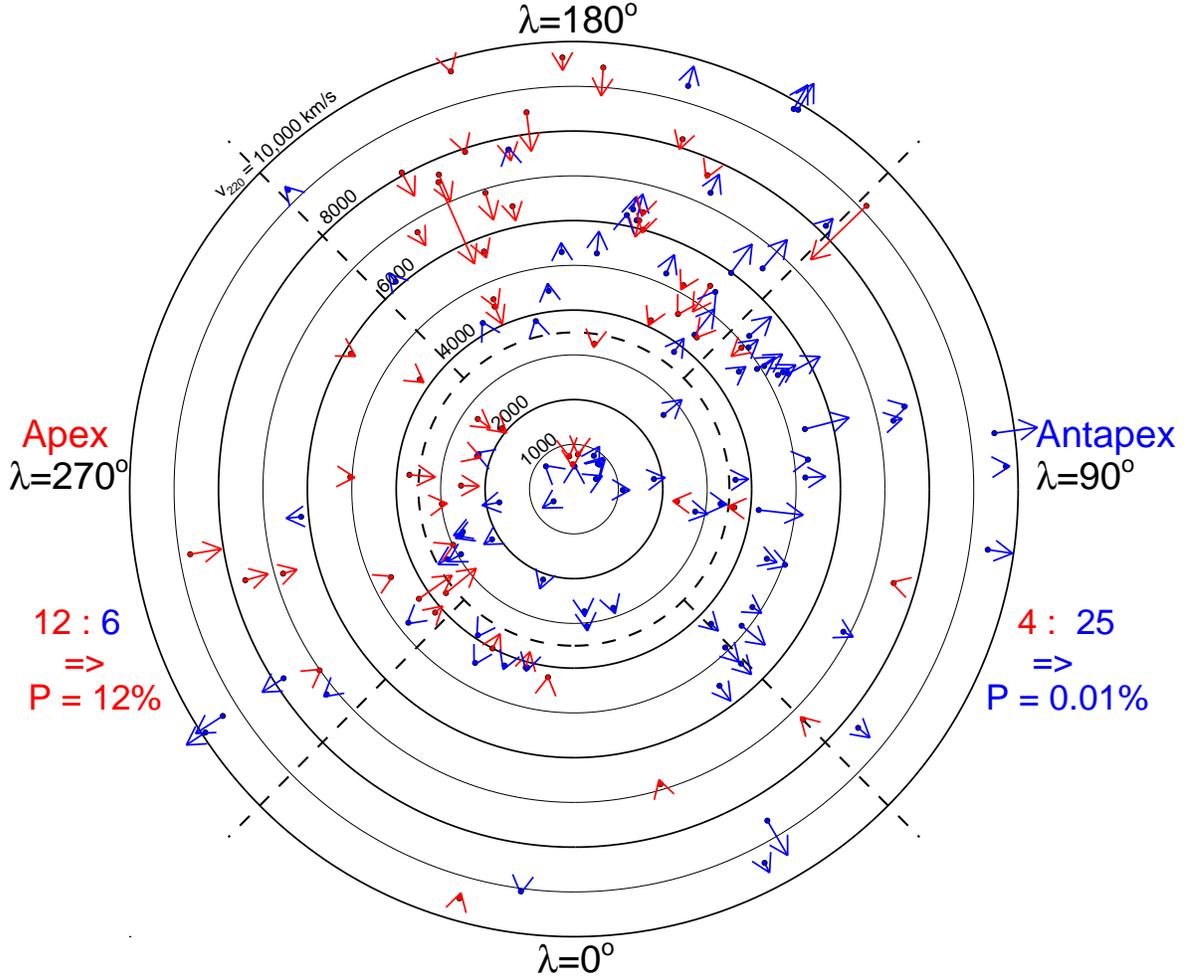}
\caption{Flow diagram of the velocity residuals 
   $\Delta v_{220}=v_{220}-v_{\rm model}$, projected on the ``apex
   plane'' (see text), of all objects with $|\beta|<45^{\circ}$ and
   $500<v_{220}<10,000\kms$. The distance scale is shown by the
   concentric circles separated by $1000\kms$. The velocity residuals
   are to the same scale. The dashed circle at $3500\kms$ denotes the
   boundary of the Local Supercluster. The bulk motion of the Local
   Supercluster causes velocity residuals that are conceived by the
   observer as ``infall'' (red arrows) and ``outflow'' (blue arrows).
   The probabilities $P$ that the distribution of red and blue arrows
   is the result of chance is shown for the $45^{\circ}$ sectors about
   the CMB apex and antapex.} 
  \label{fig:flow}
\end{figure}

     The question at which distance the cosmic value of $H_{0}$ can 
be found is answered by Figure~\ref{fig:06}: beyond $(m-M)=28.0$
($\sim\!4\;$Mpc) and out to at least $20,000\kms$ any systematic
deviation from linear expansion is limited to a few percent. At
$20,000\kms$ the expansion field is well tied to the equally linear
large-scale expansion field \citep[e.g.][]{Kessler:etal:09}. The
cosmic value of $H_{0}$ can therefore be found quite locally if
sufficient calibrators are available to compensate the locally
important peculiar velocities. The available 78 TRGB distances outside
$(m-M)=28.0$ yield a well determined value of $H_{0}=62.9\pm1.6$
(statistical error) in good agreement with local Cepheids and SNe\,Ia
\citep{TSR:08a}. Allowing for a $0.10\mag$ error of the TRGB zero
point, based on RR Lyr stars and other distance determinations, and
for a generous variation of $H_{0}$ with distance of $0.10\mag$ gives
a compounded error of $H_{0}$ of $\pm4.7$ (8\%), including the
statistical error. The discrepancy with values of $H_{0}>70$ 
\citep[e.g.][]{Riess:etal:09} is not a subject of the present paper.  

     The Local Supercluster emerges as a comoving entity of
radius $3500\kms$, corresponding to a diameter of 
$\sim\!110\;$Mpc, with the Virgo cluster at its center and several
additional clusters (UMa, Fornax, Eridanus) and a large number of groups
which must induce a network of small peculiar motions not considered
in this paper \citep[see][for a map]{Klypin:etal:03}. 
Inside the Supercluster the expansion is decelerated about inversely
to the distance from the Virgo cluster.
The galaxies of the Supercluster are concentrated toward the
supergalactic plane \citep{deVaucouleurs:56} which extends to at least
$4500\kms$ \citep{Lahav:etal:00}. It is to be noted that the CMB apex
$A_{\rm corr}$ lies at supergalactic latitude $-39^{\circ}$, which
shows that objects near to the plane (like the Great Attractor and the
Shapley Concentration) can contribute only a fraction of the
acceleration of the Local Supercluster. 

     The Supercluster's reflex motion with respect to galaxies beyond
$3500\kms$ amounts to $448\pm73\kms$ in good agreement with the
velocity of $495\pm25\kms$ toward an apex in the zone of avoidance
($l=275\pm2$, $b=12\pm4$) as inferred from the CMB dipole after
correction for the local Virgocentric infall vector. The infall vector
at the position of the LG amounts to $220\pm30\kms$ and 
diminishes for larger Virgocentric velocities according to the mass
profile of the Supercluster. The question as to the exact shape of the
Local Supercluster is very complex as evidenced for instance by the
near component of the Centaurus cluster
\citep{Lucey:etal:91,Stein:etal:97}, and the Pegasus and Hydra
clusters all three of which lie near to its outer boundary.

     The Local Supercluster is the largest volume to our knowledge for
which an unambiguous peculiar motion has been found. Claims for
coherent streaming motions over still larger scales are either
disproven or remain very doubtful in view of possible systematic
errors of the applied distance indicators with large intrinsic
scatter. In any case any large-scale streaming motion of which the
Local Supercluster could partake is limited by the agreement between
the absolute CMB dipole motion of $495\pm25\kms$ and the observed bulk
motion of $448\pm73\kms$ with respect to galaxies beyond $3500\kms$.
The question as to the relation between volume size and corresponding
peculiar velocity is decisive for the theory of structure evolution
which predicts decreasing peculiar velocities with increasing
structure size. $\Lambda$CDM models with $\Omega_{\rm M}=0.3$, 
$\Omega_{\Lambda}=0.7$ predict quite generally that a bulk motion of
$\la500\kms$ lies in the upper, yet permissible range for a volume of
radius $3500\kms$ (\citealt{Dekel:00}, Fig.~1; 
\citealt{Colless:etal:01}, Fig.~14).

     The bulk motions of galaxy concentrations outside the Local
Supercluster are still difficult to determine in view of the remaining
distance errors. The one-dimensional bulk velocities in the inertial
frame of the CMB of the 28 (super-) clusters and groups with 
$3500 < v_{\rm CMB} < 10,000\kms$ contained in Table~\ref{tab:ml} are
compiled in Table~\ref{tab:cluster} (in order of RA). 
Columns (1)$-$(3) give the name of the cluster, its adopted distance
modulus $\mu^{0}$ from SNe\,Ia, D$_{n}-\sigma$ and/or fundamental plane 
distances, and the number of distance determinations, respectively.
The observed mean cluster velocity, expressed as $v_{\rm CMB}$ is in
column~(4). The expected velocity $v_{\rm model}$ in column~(5) is
calculated for $H_{0}=62.3$, $\Omega_{\rm M}=0.3$, and 
$\Omega_{\Lambda}=0.7$ as throughout in this paper. 
The bulk velocity $v_{\rm CMB}-v_{\rm model}$ and its error follow in
column~(6). The $1\sigma$ error is estimated by assuming here a
distance error of 7.5\% for one distance determination and of 5\% for
two or three distance determinations. The average (absolute) bulk
velocity of the 28 aggregates is $353\kms$. 
However the signal of $353\kms$ is smaller than the average error of
$387\kms$, questioning whether any of the bulk velocities are
significant. If one considers instead only the 11 clusters and groups
with $3500<v_{\rm CMB}\le5000\kms$, for which the distance
errors are less effective, one obtains an average bulk motion of
$175\pm250\kms$, or $303\pm432\kms$ in three dimensions. In the sample
of 11 only the Hydra cluster has a bulk radial velocity with a
significance of $1.9\sigma$, i.e.\ $-349\pm199\kms$ toward the Local
Supercluster. Peculiar velocities of large aggregates of some hundred
$\kms$ appear to be compatible with the data. The one-dimensional bulk
velocity of the Local Supercluster of $480:\sqrt{3}=280\kms$, as seen
from a random external observer, is therefore not exeptional.

% ********************************************************
%  Table 6: One-dimensional bulk velocities of (super-)clusters
% ********************************************************

     It is unphysical to seek for a single attractor accelerating the
Local Supercluster as has been proposed in the case of the Great Attractor
(see \S~\ref{sec:2:3:1}) or the Shapley Concentration 
\citep{Mathewson:etal:92,Saunders:etal:00,Kocevski:Ebeling:06,
Basilakos:Plionis:06}, because they are 53 and 34 degrees away from
the apex direction $A_{\rm corr}$. It is clear that instead the
acceleration must be caused by the integrated force exerted by all
surrounding mass concentrations and voids. A lower distance limit of
the dominant accelerators is set by the radius of the Local
Supercluster of $\nobreak{\sim\!3500\kms}$. A direct determination of
the convergence depth is afforded by integrating the light of all-sky
samples of galaxies up to the point where the light dipole  --  which
is the same as the dipole of the gravitational pull if the
mass-to-light ratio is assumed to be constant -- agrees with the CMB dipole 
\citep[see][]{Yahil:etal:80a,Davis:Huchra:82,Lahav:87,
              Lynden-Bell:etal:89,Strauss:etal:92,Maller:etal:03}.
Particularly suited for this purpose is the relatively absorption-free
$2\;\mu$m all-sky 2MASS Redshift Survey from which 
\citet{Erdogdu:etal:06a} have derived  --  by excluding a few nearby
galaxies and integrating out to $3000\kms$  -- an apex at 
$l=269\pm9$, $b=37\pm9$. This agrees well with the CMB apex at 
$l=276\pm2$, $b=30\pm2$ as seen from the Local Group 
(see \S~\ref{sec:4}).%
% ******************************************************************
% Footnote 3
% ******************************************************************
\footnote{In this case the dipole of the integrated light
  must be compared with the CMB apex as seen from the Local Group and
  not with the apex $A_{\rm corr}$, i.e.\ corrected for Virgocentric
  infall, because the integration starts at small velocities and
  includes the Virgo cluster.} 
% ******************************************************************
The agreement is only marginally improved if the integration is carried
out to $5000\kms$. The conclusion is that most of the acceleration of
the Local Supercluster comes from a distance of $\sim\!4000\pm1000\kms$. 
A map of the dominant overdensities (in $2\;\mu$m) and voids within a
shell centered at 4000 kms-1 is given by \citet[][their
Fig.~4]{Erdogdu:etal:06b}. The structures seen in this shell are
probably the main players accelerating the Local Supercluster.

% ***********************************************
%  Acknowledgments
% ***********************************************
\acknowledgments
A.\,S. thanks the Observatories of the Carnegie Institution for
post-retirement facilities. G.\,A.\,T. thanks Prof. Norbert Straumann
for stimulating discussions. 
The authors thank the Referee for helpful comments.

% \clearpage
% ******************************************************************
% Bibliography
% ******************************************************************

% ******************************************************************

\clearpage
% ******************************************************************
% ***********              Tables                        ***********
% ******************************************************************

% ********************************************************
%  Table 1: Adopted mean distances of 281 SNe Ia and clusters
% ********************************************************
\begin{deluxetable}{lllrrrrrcrrl}
\tablewidth{0pt}
\tabletypesize{\scriptsize}
\tablecaption{{\sc Adopted Mean Distances of 281} SNe\,Ia {\sc and Clusters}\label{tab:ml}}
% ***********************************************
\tablehead{
% ***********************************************
 \colhead{Galaxy}           &  
 \colhead{SNe}              &
 \colhead{cluster}          & 
 \colhead{GLON}             &
 \colhead{GLAT}             &
 \colhead{$v_{\rm hel}$}    & 
 \colhead{$v_{220}$}        & 
 \colhead{$v_{\rm CMB}$}    & 
 \colhead{$\mu^{0}$}        &
 \colhead{Apex}             &
 \colhead{$\Delta v_{220}$} &
 \colhead{Source}           
\\
 \colhead{(1)}              &  
 \colhead{(2)}              &
 \colhead{(3)}              & 
 \colhead{(4)}              &
 \colhead{(5)}              & 
 \colhead{(6)}              & 
 \colhead{(7)}              &
 \colhead{(8)}              &
 \colhead{(9)}              &  
 \colhead{(10)}             &
 \colhead{(11)}             &            
 \colhead{(12)}                
}
% ***********************************************
\startdata
% ***********************************************
         UGC 14 &   2006sr &            & 108.87 & -38.36 &   7237 &   7352 &    7069 &  35.33 &  151 &     230 &          5 \\
         UGC 40 &   2003it &            & 110.66 & -34.33 &   7531 &   7664 &    7362 &  35.47 &  153 &      76 &          5 \\
         UGC 52 &   2002hw &            & 104.56 & -52.64 &   5257 &   5309 &    5103 &  34.66 &  138 &      53 &          5 \\
   Anon 0010-49 &   1992au &            & 319.12 & -65.88 &  18287 &  18061 &   18267 &  37.43 &   84 &    -173 &        1,3 \\
        UGC 139 &   1998dk &            & 102.85 & -62.16 &   3963 &   3981 &    3826 &  33.86 &  129 &     330 &        1,3 \\
        NGC 105 &   1997cw &            & 113.09 & -49.48 &   5290 &   5369 &    5140 &  34.09 &  139 &    1314 &        1,3 \\
   Anon 0036+11 &   1996bl &            & 116.99 & -51.30 &  10793 &  10830 &   10651 &  36.16 &  137 &     476 &      1,2,3 \\
       NGC 191A &   2006ej &            & 113.14 & -71.65 &   6131 &   6079 &    6021 &  35.00 &  119 &     -54 &          5 \\
          A2806 &  \nodata &      A2806 & 306.16 & -60.90 &   8304 &   8075 &    8310 &  35.47 &   77 &     487 &          6 \\
  MCG -02-02-86 &   2003ic &      A0085 & 115.24 & -72.03 &  16690 &  16616 &   16583 &  36.86 &  118 &    2455 &          5 \\
        NGC 232 &   2006et &            &  93.67 & -85.93 &   6647 &   6528 &    6570 &  35.04 &  106 &     283 &          5 \\
  MCG +06-02-17 &   2006mo &            & 121.85 & -26.53 &  11093 &  11245 &   10941 &  36.41 &  150 &    -339 &          5 \\
 2MASX J0056-01 &   2006gt &            & 125.75 & -64.47 &  13422 &  13382 &   13309 &  36.74 &  123 &     -43 &         4b \\
   Anon 0056-01 &   2006nz &            & 125.82 & -64.07 &  11468 &  11431 &   11355 &  36.51 &  123 &    -682 &          5 \\
        UGC 607 &   1999ef &            & 125.71 & -50.08 &  11733 &  11764 &   11602 &  36.74 &  134 &   -1660 &        1,3 \\
        UGC 646 &   1998ef &            & 125.88 & -30.57 &   5319 &   5473 &    5176 &  34.37 &  146 &     867 &        1,3 \\
        NGC 382 &   2000dk &     N383Gr & 126.84 & -30.35 &   5229 &   5380 &    5088 &  34.56 &  145 &     357 &      1,2,3 \\
     NGC 383 Gr &  \nodata &     N383Gr & 126.84 & -30.34 &   5098 &   5250 &    4957 &  34.52 &  145 &     319 &          6 \\
          A2877 &  \nodata &      A2877 & 293.13 & -70.88 &   7405 &   7194 &    7396 &  35.46 &   84 &    -359 &          6 \\
   Anon 0116+01 &   2005ir &            & 136.24 & -61.42 &  22886 &  22842 &   22782 &  38.12 &  122 &   -1840 &       4b,5 \\
     ESO 352-57 &   1992bo &            & 261.87 & -80.35 &   5549 &   5380 &    5518 &  34.87 &   92 &    -402 &      1,2,3 \\
     NGC 507 Gr &  \nodata &     N507Gr & 130.64 & -29.13 &   4934 &   5089 &    4802 &  34.35 &  142 &     525 &          6 \\
       NGC 0523 &   2001en &     N507Gr & 130.91 & -28.32 &   4758 &   4919 &    4627 &  34.27 &  142 &     518 &          5 \\
          A0194 &  \nodata &      A0194 & 142.06 & -63.10 &   5396 &   5370 &    5302 &  34.60 &  119 &     256 &        6,7 \\
  MCG -01-04-44 &   1998dm &            & 145.97 & -67.40 &   1959 &   1940 & \nodata &  33.21 &  115 &    -773 &        1,3 \\
           Anon &   2005hj &            & 142.55 & -62.76 &  17388 &  17336 &   17294 &  37.35 &  119 &    -265 &       4b,5 \\
 2MASX J0127+19 &   2005hf &      A0195 & 134.47 & -42.95 &  12924 &  12980 &   12804 &  36.64 &  134 &     141 &          5 \\
         IC 126 &   1993ae &            & 144.62 & -63.23 &   5712 &   5682 &    5621 &  34.70 &  118 &     330 &        1,3 \\
       UGC 1087 &   1999dk &            & 137.35 & -47.47 &   4485 &   4548 &    4372 &  34.22 &  130 &     246 &      1,2,3 \\
        NGC 632 &   1998es &            & 143.19 & -55.18 &   3168 &   3204 & \nodata &  33.29 &  123 &     390 &      1,2,3 \\
       UGC 1162 &   2001eh &            & 132.24 & -20.37 &  11103 &  11269 &   10976 &  36.18 &  143 &     821 &          5 \\
   Anon 0145-56 &   1992br &            & 288.01 & -59.43 &  26382 &  26119 &   26411 &  38.23 &   72 &     226 &      1,2,3 \\
        NGC 673 &   1996bo &            & 144.46 & -48.96 &   5182 &   5224 &    5084 &  34.24 &  125 &     883 &      1,2,3 \\
       UGC 1333 &   2006ob &            & 153.30 & -59.00 &  17759 &  17708 &   17680 &  37.33 &  116 &     261 &          5 \\
          A0262 &  \nodata &      A0262 & 136.59 & -25.09 &   4887 &   5048 &    4770 &  34.44 &  138 &     293 &          6 \\
  MCG +00-06-03 &   2005hc &            & 155.87 & -58.84 &  13771 &  13719 &   13696 &  36.85 &  115 &    -380 &       4b,5 \\
 2MASX J0158+36 &   2006td &            & 137.68 & -24.60 &   4761 &   4919 &    4647 &  34.65 &  137 &    -313 &          5 \\
       NGC 0809 &   2006ef &            & 169.46 & -64.80 &   5361 &   5288 &    5306 &  34.87 &  107 &    -494 &          5 \\
 MCG +06-06-12  &   2002hu &            & 141.44 & -22.27 &   8994:&   9132:&    8891:&  36.28 &  134 &   -1796:&          2 \\
   Anon 0227+28 &   2005eu &            & 147.48 & -30.06 &  10463 &  10558 &   10373 &  35.89 &  128 &    1387 &         4a \\
       UGC 1993 &   1999gp &            & 143.25 & -19.51 &   8018 &   8172 &    7922 &  35.57 &  133 &     234 &   1,2,3,4a \\
        NGC 976 &   1999dq &            & 152.84 & -35.87 &   4295 &   4387 &    4218 &  33.79 &  123 &     851 &      1,2,3 \\
        IC 1844 &   1995ak &            & 169.66 & -48.98 &   6811 &   6785 &    6767 &  35.04 &  109 &     540 &      1,2,3 \\
       UGC 2320 &   2003iv &            & 162.50 & -40.73 &  10285 &  10293 &   10230 &  36.29 &  114 &    -684 &          5 \\
   CGCG 539-121 &   2005ls &            & 145.75 & -14.64 &   6331 &   6508 &    6245 &  34.82 &  130 &     856 &          5 \\
       UGC 2384 &   2006os &      A0397 & 161.38 & -37.47 &   9836 &   9864 &    9780 &  35.86 &  116 &     817 &          5 \\
          A0397 &  \nodata &      A0397 & 161.91 & -37.24 &   9803 &   9830 &    9749 &  35.87 &  115 &     742 &          6 \\
          A0400 &  \nodata &      A0400 & 170.24 & -44.93 &   7315 &   7299 &    7276 &  35.19 &  108 &     615 &          6 \\
     ESO 300-09 &   1992bc &            & 245.70 & -59.64 &   5996 &   5791 &    6036 &  35.03 &   75 &    -426 &      1,2,3 \\
  MCG -01-09-06 &   2005eq &            & 187.64 & -51.74 &   8687 &   8598 &    8677 &  35.78 &   97 &    -129 &       4a,b \\
       NGC 1259 &    2008L &      A0426 & 150.24 & -13.62 &   5816 &   5988 &    5745 &  34.63 &  126 &     803 &          5 \\
       NGC 1275 &   2005mz &      A0426 & 150.58 & -13.26 &   5264 &   5435 &    5194 &  34.72 &  125 &      34 &          5 \\
       NGC 1316 &    1981D &     Fornax & 240.16 & -56.69 &   1760 &   1371 & \nodata &  31.23 &   74 & \nodata &        1,3 \\
       NGC 1316 &    1980N &     Fornax & 240.16 & -56.69 &   1760 &   1371 & \nodata &  31.71 &   74 & \nodata &        1,3 \\
   Anon 0329-37 &   1992bs &            & 240.03 & -55.34 &  18887 &  18659 &   18938 &  37.68 &   73 &   -1699 &      1,2,3 \\
       NGC 1380 &    1992A &     Fornax & 235.93 & -54.06 &   1877 &   1371 & \nodata &  31.82 &   74 &     -64 &      1,2,3 \\
   Anon 0336-18 &   1992bp &            & 208.83 & -51.09 &  23684 &  23525 &   23713 &  37.79 &   85 &    2160 &      1,2,3 \\
   Anon 0335-33 &    1990Y &            & 232.64 & -53.85 &  11702 &  11496 &   11753 &  36.38 &   75 &      68 &        1,3 \\
       UGC 2829 &   2006kf &            & 178.55 & -35.71 &   6386 &   6378 &    6378 &  35.12 &  102 &     -98 &          5 \\
     ESO 156-08 &   1992bk &      A3158 & 265.05 & -48.93 &  17598 &  17333 &   17675 &  37.27 &   61 &     343 &      1,2,3 \\
       NGC 1448 &   2001el &            & 251.52 & -51.39 &   1168 &   1015 & \nodata &  31.73 &   66 &    -362 &      1,2,3 \\
          Grm13 &  \nodata &      Grm13 & 266.04 & -43.47 &    929 &    754 & \nodata &  31.74 &   56 &    -629 &          7 \\
          A0496 &  \nodata &      A0496 & 209.57 & -36.48 &   9863 &   9736 &    9928 &  36.03 &   78 &     -31 &          6 \\
       UGC 3108 &   2006lf &            & 159.92 &  -1.93 &   3959 &   4150 &    3930 &  34.21 &  115 &    -132 &         4a \\
       NGC 1699 &   2001ep &            & 203.60 & -27.55 &   3901 &   3850 &    3971 &  33.95 &   79 &      46 &          5 \\
   Anon 0459-58 &   1992bh &            & 267.85 & -37.33 &  13491 &  13220 &   13602 &  36.88 &   50 &   -1069 &      1,2,3 \\
       UGC 3218 &   2006le &            & 147.89 &  12.18 &   5226 &   5479 &    5176 &  35.03 &  122 &    -738 &         4a \\
       NGC 1819 &   2005el &            & 196.17 & -19.40 &   4470 &   4466 &    4535 &  34.21 &   83 &     183 &       4a,b \\
          A0539 &  \nodata &      A0539 & 195.70 & -17.72 &   8514 &   8486 &    8580 &  35.95 &   83 &    -936 &          7 \\
       UGC 3329 &   1999ek &            & 189.40 &  -8.23 &   5253 &   5304 &    5314 &  34.59 &   87 &     213 &   1,2,3,4a \\
      PGC 17787 &   1993ac &            & 149.72 &  17.21 &  14690 &  14923 &   14652 &  37.00 &  118 &    -149 &      1,2,3 \\
    CGCG 308-09 &    2006N &            & 149.44 &  19.97 &   4280 &   4561 &    4245 &  34.36 &  117 &     -24 &         4a \\
          A3381 &  \nodata &      A3381 & 240.29 & -22.70 &  10763 &  10561 &   10907 &  36.32 &   48 &    -565 &          7 \\
       UGC 3432 &   1996bv &            & 157.34 &  17.97 &   4998 &   5247 &    4984 &  34.44 &  111 &     491 &      1,2,3 \\
 2MASX J0627-35 &   1999ao &            & 243.83 & -20.02 &  16189 &  15974 &   16342 &  37.21 &   44 &    -570 &          5 \\
   CGCG 233-023 &   2002kf &            & 165.59 &  18.25 &   5786 &   5999 &    5799 &  34.98 &  104 &     -79 &          5 \\
     ESO 427-06 &    2004S &            & 240.79 & -14.79 &   2806 &   2672 & \nodata &  33.47 &   43 &    -383 &          2 \\
       NGC 2258 &    1997E &            & 140.22 &  25.82 &   4059 &   4374 &    4007 &  34.26 &  122 &      -7 &      1,2,3 \\
       UGC 3576 &   1998ec &            & 166.30 &  20.71 &   5966 &   6180 &    5984 &  35.16 &  102 &    -415 &        1,3 \\
       UGC 3634 &   2005na &            & 201.40 &   8.61 &   7891 &   7923 &    8005 &  35.57 &   72 &     -16 &       4a,b \\
       NGC 2320 &    2000B &            & 166.36 &  22.79 &   5944 &   6169 &    5964 &  34.76 &  102 &     669 &        1,3 \\
          A0569 &  \nodata &      A0569 & 168.57 &  22.81 &   6026 &   6237 &    6053 &  35.03 &  100 &      20 &          6 \\
       UGC 3725 &   2007au &      A0569 & 167.38 &  23.53 &   6171 &   6390 &    6195 &  34.98 &  101 &     312 &          5 \\
       NGC 2268 &    1982B &            & 129.24 &  27.55 &   2222 &   2610 & \nodata &  32.48 &  128 &     668 &          1 \\
       UGC 3770 &   2000fa &            & 194.17 &  15.48 &   6378 &   6469 &    6477 &  35.23 &   78 &    -338 &      1,2,3 \\
       UGC 3787 &   2003ch &            & 207.22 &  10.26 &   7495 &   7504 &    7626 &  35.97 &   66 &   -2002 &          5 \\
       UGC 3845 &   1997do &            & 171.00 &  25.27 &   3034 &   3279 & \nodata &  33.69 &   97 &     -99 &      1,2,3 \\
   Anon 0741-62 &   1992bg &            & 274.61 & -18.35 &  10793 &  10537 &   10947 &  36.28 &   30 &    -391 &      1,2,3 \\
  MCG +08-14-43 &    2007R &            & 174.35 &  28.16 &   9258 &   9443 &    9307 &  36.12 &   93 &    -727 &          5 \\
       NGC 2441 &    1995E &            & 141.99 &  30.26 &   3470 &   3804 &    3430 &  33.66 &  118 &     472 &      1,2,3 \\
       UGC 4133 &   2006qo &            & 161.38 &  31.68 &   9130 &   9368 &    9145 &  35.80 &  103 &     561 &          5 \\
       UGC 4195 &   2000ce &            & 149.10 &  32.00 &   4888 &   5180 &    4869 &  34.69 &  112 &    -148 &   1,2,3,4a \\
    CGCG 207-42 &   2006te &            & 178.98 &  32.08 &   9471 &   9648 &    9536 &  36.08 &   88 &    -341 &          5 \\
          A0634 &  \nodata &      A0634 & 159.40 &  33.64 &   7945 &   8192 &    7957 &  35.72 &  104 &    -303 &          6 \\
       UGC 4322 &   2002he &            & 153.60 &  33.98 &   7364 &   7631 &    7360 &  35.51 &  108 &     -95 &          5 \\
         Cancer &  \nodata &     Cancer & 202.55 &  28.69 &   4497 &   4613 &    4623 &  34.49 &   69 &    -252 &          6 \\
       NGC 4414 &    1974G &            & 174.54 &  83.18 &    716 &    788 & \nodata &  31.47 &   79 &    -433 &        1,2 \\
       UGC 4414 &   2005mc &            & 202.50 &  30.23 &   7561 &   7656 &    7687 &  35.69 &   69 &    -724 &          5 \\
       NGC 2595 &   1999aa &     Cancer & 202.73 &  30.31 &   4330 &   4452 &    4457 &  34.52 &   69 &    -479 &      1,2,3 \\
       UGC 4455 &   2007bd &            & 226.07 &  21.52 &   9299 &   9277 &    9474 &  35.89 &   47 &     107 &          5 \\
       NGC 2623 &   1999gd &            & 198.84 &  33.97 &   5549 &   5682 &    5666 &  34.90 &   72 &    -179 &      1,2,3 \\
  MCG +03-22-20 &   2004gs &            & 207.96 &  31.32 &   7988 &   8065 &    8126 &  35.78 &   64 &    -662 &         4b \\
       UGC 4614 &   2005ms &            & 186.90 &  38.49 &   7556 &   7727 &    7644 &  35.60 &   81 &    -319 &          5 \\
  MCG -01-23-08 &   2002hd &            & 234.77 &  23.16 &  10493 &  10449 &   10680 &  35.93 &   40 &    1112 &          5 \\
    CGCG 180-22 &    1999X &            & 186.59 &  39.59 &   7546 &   7726 &    7634 &  35.49 &   81 &      70 &        1,3 \\
  MCG +08-17-43 &    2001G &            & 168.32 &  42.31 &   5028 &   5292 &    5072 &  34.37 &   94 &     686 &          5 \\
          A0779 &  \nodata &      A0779 & 191.07 &  44.41 &   6742 &   6913 &    6840 &  35.52 &   77 &    -848 &          6 \\
       NGC 2935 &    1996Z &            & 253.59 &  22.57 &   2271 &   2280 & \nodata &  32.76 &   23 &      72 &      1,2,3 \\
       NGC 2930 &    2005M &            & 206.89 &  46.22 &   6599 &   6733 &    6728 &  35.35 &   66 &    -454 &         4b \\
       UGC 5129 &   2001fe &            & 203.70 &  46.88 &   4059 &   4240 &    4182 &  34.17 &   68 &      35 &          5 \\
       NGC 2962 &    1995D &            & 230.00 &  39.67 &   1966 &   2120 & \nodata &  32.85 &   48 &    -180 &      1,2,3 \\
       NGC 2986 &   1999gh &            & 255.04 &  23.72 &   2302 &   2300 & \nodata &  32.94 &   22 &     -97 &      1,2,3 \\
       UGC 5234 &    2003W &            & 217.68 &  45.93 &   6017 &   6132 &    6164 &  34.89 &   59 &     297 &          5 \\
       NGC 3021 &   1995al &            & 192.18 &  50.84 &   1541 &   1841 & \nodata &  32.71 &   76 &    -317 &      1,2,3 \\
       UGC 5378 &    2007S &            & 234.37 &  43.38 &   4161 &   4246 &    4332 &  34.26 &   47 &    -135 &          5 \\
       UGC 5542 &   2001ie &            & 150.36 &  47.78 &   9215 &   9490 &    9223 &  35.89 &  103 &     319 &          5 \\
   Anon 1003-35 &   1993ag &            & 268.44 &  15.93 &  14700 &  14542 &   14898 &  37.03 &    7 &    -732 &      1,2,3 \\
   Anon 1009-26 &    1992J &            & 263.55 &  23.54 &  13491 &  13375 &   13691 &  36.62 &   16 &     651 &        1,3 \\
    CGCG 266-31 &   2002bf &            & 156.46 &  50.08 &   7254 &   7521 &    7278 &  35.60 &   98 &    -525 &          3 \\
       NGC 3147 &   1997bq &            & 136.29 &  39.46 &   2820 &   3188 & \nodata &  33.45 &  116 &     161 &      1,2,3 \\
       NGC 3190 &   2002bo &            & 213.04 &  54.85 &   1271 &   1573 & \nodata &  32.20 &   64 &    -135 &   1,2,3,4a \\
  MCG +03-27-38 &    2004L &            & 223.79 &  54.79 &   9686 &   9790 &    9829 &  36.08 &   58 &    -199 &          5 \\
       UGC 5691 &    1991S &            & 214.07 &  57.42 &  16489 &  16610 &   16617 &  37.28 &   64 &    -455 &        1,3 \\
          AS636 &  \nodata &     Antlia & 272.95 &  19.19 &   2608 &   2541 & \nodata &  33.40 &    8 &    -418 &          6 \\
   Anon 1034-34 &    1993B &            & 273.33 &  20.46 &  20686 &  20538 &   20881 &  37.75 &    9 &    -456 &      1,2,3 \\
       NGC 3294 &    1992G &            & 184.62 & -59.84 &   1586 &   1542 & \nodata &  32.70 &  100 &    -606 &          3 \\
          A1060 &  \nodata &      Hydra & 269.63 &  26.51 &   3777 &   3717 &    3973 &  34.23 &   15 &    -604 &        6,7 \\
  MCG +11-13-36 &   2006ar &            & 142.76 &  46.62 &   6757 &   7054 &    6747 &  35.46 &  107 &    -499 &          5 \\
   Anon 1039+05 &   2006al &      A1066 & 241.92 &  51.69 &  20341 &  20380 &   20505 &  37.72 &   48 &    -339 &          5 \\
       NGC 3327 &    2001N &            & 211.35 &  60.29 &   6303 &   6471 &    6424 &  35.11 &   67 &      24 &          5 \\
       NGC 3332 &   2005ki &            & 236.83 &  54.27 &   5758 &   5858 &    5914 &  34.96 &   52 &    -165 &       4b,5 \\
          AS639 &  \nodata &      AS639 & 280.54 &  10.91 &   6326 &   6171 &    6512 &  34.89 &    6 &     337 &          7 \\
       NGC 3368 &   1998bu &            & 234.44 &  57.01 &    897 &    719 & \nodata &  30.45 &   55 &     -45 &   1,2,3,4a \\
       NGC 3370 &   1994ae &            & 225.35 &  59.67 &   1279 &   1603 & \nodata &  32.63 &   60 &    -477 &      1,2,3 \\
       UGC 6015 &   2006cf &            & 165.96 &  60.09 &  12457 &  12690 &   12509 &  36.55 &   89 &     357 &          5 \\
   Anon 1101-06 &   1999aw &            & 260.24 &  47.45 &  11992 &  11994 &   12168 &  36.59 &   38 &    -560 &      1,2,3 \\
 2MASX J1109+28 &   2006ak &            & 203.15 &  67.48 &  11422 &  11595 &   11523 &  36.42 &   72 &     -41 &          5 \\
    NGC 3557 Gr &  \nodata &    N3557Gr & 281.58 &  21.09 &   3056 &   2990 & \nodata &  33.51 &   11 &    -121 &          6 \\
       UGC 6211 &   2001ah &            & 149.13 &  56.52 &  17315 &  17573 &   17334 &  37.21 &   98 &    1028 &          5 \\
       UGC 6332 &   2007bc &            & 224.67 &  68.06 &   6227 &   6394 &    6346 &  35.13 &   65 &    -111 &          5 \\
       NGC 3627 &    1989B &            & 241.96 &  64.42 &    727 &    431 & \nodata &  30.35 &   57 &    -299 &      1,2,3 \\
       UGC 6363 &   2004bg &            & 223.72 &  68.62 &   6306 &   6477 &    6423 &  35.10 &   65 &      60 &          5 \\
       NGC 3663 &   2006ax &            & 271.83 &  45.23 &   5018 &   5041 &    5194 &  34.62 &   34 &    -120 &       4a,b \\
      HOLM 254B &   2004as &            & 220.68 &  70.12 &   9300 &   9457 &    9412 &  36.13 &   67 &    -759 &          5 \\
          A1314 &  \nodata &      A1314 & 151.83 &  63.57 &  10043 &  10302 &   10076 &  36.05 &   93 &     446 &          6 \\
     ESO 439-18 &   2001ba &            & 285.38 &  28.03 &   8861 &   8759 &    9041 &  35.97 &   19 &    -748 &     2,3,4a \\
  EROS J1139-08 &   1999bp &            & 274.68 &  50.04 &  23084 &  23075 &   23250 &  38.06 &   38 &    -970 &          5 \\
          A1367 &  \nodata &      A1367 & 234.80 &  73.03 &   6595 &   6765 &    6707 &  35.17 &   65 &     140 &          6 \\
       NGC 3873 &   2007ci &      A1367 & 235.50 &  73.26 &   5434 &   5618 &    5546 &  34.78 &   66 &      68 &          5 \\
       NGC 3978 &   2003cq &            & 134.85 &  55.31 &   9978 &  10270 &    9971 &  36.07 &  105 &     326 &          5 \\
       NGC 3982 &   1998aq &            & 138.83 &  60.27 &   1109 &   1517 & \nodata &  32.07 &  100 &     -92 &      1,2,3 \\
       NGC 3987 &    2001V &            & 218.95 &  77.72 &   4502 &   4752 &    4596 &  34.18 &   72 &     527 &      1,2,3 \\
       NGC 4172 &   2006az &            & 133.89 &  60.11 &   9274 &   9563 &    9276 &  35.91 &  102 &     309 &          5 \\
       UGC 7357 &   2006cp &            & 243.90 &  81.31 &   6682 &   6873 &    6773 &  35.31 &   71 &    -185 &         4a \\
     ESO 573-14 &   2000bh &            & 293.74 &  40.34 &   6838 &   6812 &    6997 &  35.35 &   33 &    -375 &     2,3,4a \\
       NGC 4321 &    2006X &      Virgo & 271.14 &  76.90 &   1571 &   1152 & \nodata &  31.42 &   65 &     -42 &         4a \\
       NGC 4419 &    1984A &      Virgo & 276.45 &  76.64 &   -261 &   1152 & \nodata &  31.20 &   65 & \nodata &          1 \\
       NGC 4493 &    1994M &            & 291.69 &  63.04 &   6943 &   7034 &    7072 &  35.39 &   53 &    -284 &      1,2,3 \\
       NGC 4495 &    1994S &            & 187.34 &  85.14 &   4550 &   4804 &    4619 &  34.47 &   78 &     -17 &      1,2,3 \\
      NGC 4496A &    1960F &       VirW & 290.56 &  66.33 &   1730 &   1168 & \nodata &  30.81 &   56 &     266 &          1 \\
       NGC 4501 &   1999cl &      Virgo & 282.33 &  76.51 &   2281 &   1152 & \nodata &  31.33 &   65 & \nodata &         4a \\
       NGC 4520 &   2000bk &            & 295.26 &  55.23 &   7628 &   7671 &    7767 &  35.70 &   46 &    -747 &     1,3,4a \\
       NGC 4526 &    1994D &      Virgo & 290.16 &  70.14 &    448 &   1152 & \nodata &  31.36 &   59 & \nodata &      1,2,3 \\
       NGC 4536 &    1981B &       VirW & 292.95 &  64.73 &   1808 &   1407 & \nodata &  31.26 &   54 & \nodata &      1,2,3 \\
       NGC 4619 &   2006ac &            & 136.98 &  81.80 &   6927 &   7174 &    6976 &  35.31 &   84 &     116 &         4a \\
          Cen30 &  \nodata &      Cen30 & 300.97 &  22.15 &   3041 &   2991 & \nodata &  33.44 &   27 &     -22 &          6 \\
        IC 3690 &    1992P &            & 295.62 &  73.11 &   7615 &   7751 &    7720 &  35.71 &   63 &    -705 &      1,2,3 \\
       NGC 4639 &    1990N &      Virgo & 294.29 &  75.99 &   1018 &   1152 & \nodata &  32.16 &   65 & \nodata &      1,2,3 \\
       NGC 4675 &    1997Y &            & 124.77 &  62.37 &   4757 &   5083 &    4753 &  34.60 &  102 &     -32 &      1,2,3 \\
       UGC 7934 &    2006S &            & 131.43 &  81.95 &   9624 &   9856 &    9671 &  36.15 &   85 &    -453 &          5 \\
 2MASX J1246+12 &   2004gu &            & 298.25 &  74.78 &  13748 &  13875 &   13848 &  36.78 &   64 &     209 &         4b \\
       NGC 4680 &   1997bp &            & 301.16 &  51.22 &   2492 &   2675 & \nodata &  32.92 &   45 &     299 &      1,2,3 \\
       NGC 4679 &   2001cz &            & 302.11 &  23.29 &   4643 &   4563 &    4794 &  34.37 &   29 &     -43 &     2,3,4a \\
       NGC 4704 &   1998ab &            & 124.87 &  75.20 &   8134 &   8407 &    8162 &  35.37 &   91 &    1154 &      1,2,3 \\
       NGC 4753 &    1983G &            & 303.42 &  61.67 &   1239 &   1264 & \nodata &  31.45 &   54 &      53 &          1 \\
  MGC -01-33-34 &    2006D &            & 303.40 &  53.09 &   2556 &   2736 & \nodata &  33.06 &   47 &     203 &         4a \\
 2MASX J1259+28 &   2006cj &            &  68.06 &  87.86 &  20241 &  20433 &   20297 &  37.96 &   80 &   -2583 &          5 \\
          A1656 &  \nodata &       Coma &  58.08 &  87.96 &   6925 &   7152 &    6982 &  35.37 &   80 &    -100 &        6,7 \\
       IC 4042A &   2006bz &       Coma &  55.57 &  87.78 &   8366 &   8583 &    8423 &  35.73 &   80 &      50 &          5 \\
       UGC 8162 &    2007F &            & 118.24 &  66.40 &   7072 &   7367 &    7072 &  35.44 &  100 &    -118 &          5 \\
 2MASX J1305+28 &   2006cg &            &  61.91 &  86.59 &   8413 &   8645 &    8466 &  35.26 &   81 &    1745 &          5 \\
        IC 4182 &    1937C &            & 107.70 &  79.09 &    321 &    298 & \nodata &  28.51 &   89 &     -15 &        1,2 \\
  MCG +06-29-43 &    2002G &            &  96.96 &  82.19 &  10114 &  10347 &   10152 &  36.21 &   86 &    -243 &          5 \\
     ESO 508 Gr &  \nodata &    E508 Gr & 307.98 &  39.08 &   3196 &   3252 & \nodata &  33.41 &   40 &     279 &          6 \\
       NGC 5018 &   2002dj &            & 309.90 &  43.06 &   2816 &   2928 & \nodata &  33.05 &   44 &     407 &          5 \\
       NGC 5061 &    1996X &            & 310.25 &  35.66 &   2065 &   2157 & \nodata &  32.45 &   40 &     242 &      1,2,3 \\
      PGC 46640 &    1994T &            & 318.02 &  59.84 &  10390 &  10465 &   10493 &  36.20 &   58 &     -78 &      1,2,3 \\
        IC 4232 &    1991U &            & 311.82 &  36.21 &   9426 &   9389 &    9554 &  35.74 &   41 &     818 &        1,3 \\
     ESO 508-67 &   1992ag &            & 312.49 &  38.39 &   7795 &   7783 &    7921 &  35.33 &   43 &     661 &      1,2,3 \\
        IC 4239 &   2006cq &            &  62.88 &  81.84 &  14491 &  14707 &   14528 &  36.98 &   85 &    -231 &          5 \\
     ESO 508-75 &   2007cg &            & 312.72 &  37.57 &   9952 &   9919 &   10078 &  35.93 &   43 &     582 &          5 \\
  MCG -02-34-61 &   2007ca &            & 316.95 &  46.69 &   4217 &   4264 &    4331 &  34.73 &   50 &   -1162 &          5 \\
   Anon 1331-33 &    1993O &            & 312.42 &  28.92 &  15589 &  15502 &   15717 &  37.18 &   39 &    -824 &      1,2,3 \\
     ESO 383-32 &   2000ca &            & 313.20 &  27.83 &   7080 &   7013 &    7206 &  35.31 &   39 &     -45 &     2,3,4a \\
       NGC 5253 &    1972E &            & 314.86 &  30.11 &    407 &    171 & \nodata &  27.94 &   42 &     -69 &      1,2,3 \\
       NGC 5283 &   2005dv &            & 115.81 &  48.76 &   3119 &   3489 & \nodata &  33.93 &  117 &    -281 &          5 \\
       NGC 5308 &   1996bk &            & 111.25 &  54.88 &   2041 &   2459 & \nodata &  32.62 &  112 &     389 &      1,2,3 \\
          A3574 &  \nodata &      A3574 & 317.46 &  30.94 &   4797 &   4775 &    4913 &  34.43 &   44 &      41 &        6,7 \\
       NGC 5304 &   2005al &      A3574 & 317.59 &  30.62 &   3718 &   3699 &    3833 &  34.39 &   44 &    -949 &         4b \\
  MCG +08-25-47 &    1996C &            &  99.62 &  65.04 &   8094 &   8384 &    8077 &  36.05 &  103 &   -1472 &      1,2,3 \\
     ESO 445-66 &    1993H &            & 318.22 &  30.33 &   7257 &   7211 &    7371 &  35.33 &   44 &      89 &      1,2,3 \\
          AS753 &  \nodata &      AS753 & 319.63 &  26.55 &   4197 &   4167 &    4306 &  34.27 &   45 &    -234 &          7 \\
       NGC 5468 &   1999cp &            & 334.87 &  52.70 &   2842 &   3010 & \nodata &  33.46 &   63 &     -32 &     2,3,4a \\
       NGC 5468 &   2002cr &            & 334.87 &  52.70 &   2842 &   3006 & \nodata &  33.58 &   63 &    -207 &          5 \\
  MCG +05-34-33 &   2002bz &            &  36.89 &  69.23 &  11138 &  11357 &   11146 &  36.47 &   90 &    -542 &          5 \\
        IC 4423 &   2001ay &            &  35.97 &  68.82 &   9067 &   9290 &    9075 &  36.09 &   90 &    -744 &        2,3 \\
   Anon 1433+03 &   2006bw &            & 353.79 &  56.19 &   8994 &   9123 &    9036 &  35.87 &   74 &      35 &          5 \\
       UGC 9391 &   2003du &            & 101.18 &  53.21 &   1914 &   2306 & \nodata &  33.24 &  115 &    -445 &     2,3,4a \\
       UGC 9612 &    2007O &            &  77.66 &  59.25 &  10856 &  11144 &   10813 &  36.16 &  108 &     789 &          5 \\
           Anon &   2005ag &            &   7.85 &  55.50 &  23807 &  23945 &   23822 &  37.94 &   82 &    1128 &         4b \\
       UGC 9640 &   2008af &            &  20.11 &  58.62 &  10045 &  10236 &   10046 &  35.99 &   88 &     643 &          5 \\
  MCG -01-39-03 &   2005cf &            & 354.81 &  39.85 &   1937 &   2136 & \nodata &  32.38 &   75 &     282 &          4 \\
      UGC 10030 &   2002ck &            &   6.55 &  39.30 &   8953 &   9068 &    8946 &  35.85 &   84 &      61 &          5 \\
  MCG +11-19-25 &   2000cf &            &  99.88 &  42.17 &  10920 &  11241 &   10831 &  36.42 &  126 &    -395 &      1,2,3 \\
  MCG +03-41-03 &   2007ap &            &  28.86 &  46.04 &   4742 &   4964 &    4701 &  34.77 &   98 &    -561 &          5 \\
    CGCG 108-13 &   2006bt &            &  33.70 &  47.24 &   9640 &   9850 &    9593 &  36.03 &  100 &      83 &          5 \\
        IC 1151 &    1991M &            &  30.36 &  45.90 &   2169 &   2436 & \nodata &  33.64 &   99 &    -865 &          3 \\
       NGC 6038 &   1999cc &            &  59.67 &  48.75 &   9392 &   9665 &    9320 &  36.01 &  112 &     -15 &      1,2,3 \\
       NGC 6063 &   1999ac &            &  19.89 &  39.95 &   2848 &   3077 & \nodata &  33.38 &   94 &     145 &      1,2,3 \\
      UGC 10244 &   2006cc &            &  68.18 &  47.10 &   9752 &  10037 &    9670 &  36.31 &  117 &   -1038 &          5 \\
     ESO 584-07 &   2007ai &            & 353.04 &  21.09 &   9492 &   9500 &    9499 &  36.00 &   75 &    -137 &          5 \\
       NGC 6104 &   2002de &            &  57.37 &  45.91 &   8429 &   8702 &    8350 &  35.74 &  113 &     131 &          5 \\
          A2199 &  \nodata &      A2199 &  62.70 &  43.70 &   9039 &   9322 &    8949 &  35.76 &  117 &     673 &          6 \\
      UGC 10483 &   2001az &            & 108.95 &  34.29 &  12200 &  12520 &   12100 &  36.58 &  132 &      21 &          5 \\
      PGC 59076 &    1994Q &      A2199 &  64.39 &  39.68 &   8863 &   9148 &    8760 &  35.87 &  121 &      59 &        1,3 \\
      UGC 10704 &   2007ae &            & 111.49 &  31.71 &  19303 &  19615 &   19201 &  37.33 &  134 &    2168 &          5 \\
      UGC 10738 &   2001cp &            &  26.50 &  24.95 &   6716 &   6874 &    6634 &  35.29 &  104 &    -120 &          5 \\
      UGC 10743 &   2002er &            &  28.67 &  25.83 &   2569 &   2796 & \nodata &  33.15 &  106 &     157 &      1,2,3 \\
  MCG +03-44-03 &    1990O &            &  37.65 &  28.37 &   9193 &   9389 &    9095 &  35.87 &  112 &     301 &      1,2,3 \\
      NGC 6365A &    2003U &            &  91.47 &  34.06 &   8496 &   8823 &    8377 &  35.64 &  134 &     630 &          5 \\
       NGC 6462 &   2005ao &            &  90.99 &  31.49 &  11514 &  11828 &   11388 &  36.52 &  137 &    -340 &         4a \\
      UGC 11064 &   2000cn &            &  53.45 &  23.32 &   7043 &   7285 &    6908 &  35.41 &  127 &     -99 &      1,2,3 \\
  MCG +04-42-22 &   2001bf &            &  52.16 &  21.97 &   4647 &   4911 &    4511 &  34.26 &  126 &     530 &          5 \\
      UGC 11149 &   1998dx &            &  77.68 &  26.67 &  16256 &  16546 &   16113 &  37.02 &  138 &    1339 &      1,2,3 \\
       NGC 6627 &    1998V &            &  43.94 &  13.34 &   5272 &   5480 &    5132 &  34.56 &  124 &     458 &      1,2,3 \\
  MCG +05-43-16 &   2007co &            &  57.59 &  18.82 &   8083 &   8329 &    7933 &  35.57 &  132 &     391 &          5 \\
       NGC 6685 &   2006bq &            &  68.85 &  19.09 &   6567 &   6851 &    6409 &  35.15 &  140 &     286 &          5 \\
        IC 4758 &   2001cn &     PavoII & 329.65 & -24.05 &   4647 &   4475 &    4676 &  34.28 &   65 &      54 &     2,3,4a \\
          AS805 &  \nodata &     PavoII & 332.25 & -23.59 &   4167 &   4002 &    4189 &  34.28 &   67 &    -419 &          6 \\
 2MASX J1911+77 &   2003hu &            & 109.56 &  25.44 &  22484 &  22790 &   22364 &  37.61 &  140 &    3050 &          5 \\
        IC 4830 &   2001bt &            & 337.32 & -25.87 &   4388 &   4240 &    4392 &  34.13 &   72 &     111 &     2,3,4a \\
  MCG +07-41-01 &   2002do &            &  75.61 &   6.13 &   4761 &   5041 &    4578 &  34.54 &  154 &      65 &          5 \\
      PGC 63925 &    1990T &            & 341.50 & -31.52 &  11992 &  11809 &   11977 &  36.63 &   77 &    -972 &        1,3 \\
        IC 4919 &   1991ag &      Grm15 & 342.55 & -31.64 &   4264 &   4124 &    4246 &  34.07 &   78 &     106 &      1,2,3 \\
          Grm15 &  \nodata &      Grm15 & 341.85 & -32.45 &   4286 &   4134 &    4269 &  34.36 &   78 &    -451 &          7 \\
          PavoI &  \nodata &      PavoI & 324.10 & -32.58 &   4107 &   3916 &    4138 &  33.98 &   65 &      60 &          6 \\
       NGC 6928 &   2004eo &            &  54.16 & -17.26 &   4707 &   4855 &    4521 &  34.15 &  141 &     688 &       4a,b \\
       NGC 6951 &    2000E &            & 100.90 &  14.85 &   1424 &   1814 & \nodata &  31.96 &  153 &     284 &      1,2,3 \\
     ESO 234-69 &   1992al &            & 347.34 & -38.49 &   4381 &   4242 &    4343 &  34.19 &   84 &      -2 &      1,2,3 \\
       NGC 6962 &   2002ha &            &  47.41 & -25.37 &   4211 &   4311 &    4035 &  34.10 &  134 &     238 &          5 \\
       NGC 6986 &   2002el &            &  28.77 & -35.67 &   8612 &   8591 &    8473 &  35.46 &  116 &    1038 &          2 \\
 2MASX J2120+44 &   2001fh &            &  88.22 &  -3.81 &   3894 &   4185 &    3702 &  33.55 &  170 &    1016 &          5 \\
   Anon 2123-00 &   2006oa &            &  51.74 & -33.74 &  17988 &  18027 &   17810 &  37.35 &  135 &     425 &          5 \\
   Anon 2128-61 &   1992ae &            & 332.70 & -41.99 &  22484 &  22260 &   22479 &  37.76 &   76 &    1174 &      1,2,3 \\
   Anon 2135-62 &   1990af &            & 330.82 & -42.24 &  15080 &  14856 &   15079 &  36.90 &   75 &     440 &      1,2,3 \\
       NGC 7131 &   1998co &            &  41.52 & -44.94 &   5418 &   5419 &    5264 &  34.55 &  124 &     419 &          3 \\
      UGC 11816 &   2004ey &            &  57.47 & -38.27 &   4733 &   4804 &    4553 &  34.33 &  138 &     282 &         4b \\
   Anon 2155-01 &   2006on &            &  57.18 & -40.55 &  20985 &  21009 &   20808 &  37.68 &  136 &     651 &          5 \\
          MDL59 &  \nodata &      MDL59 &  14.95 &  24.39 &   2567 &   2740 & \nodata &  33.12 &   94 &     137 &          6 \\
        IC 5179 &   1999ee &            &   6.50 & -55.93 &   3422 &   3325 & \nodata &  33.57 &  101 &     127 &   1,2,3,4a \\
      UGC 12071 &   2006gr &            &  90.75 &  38.36 &  10372 &  10689 &   10265 &  36.39 &  130 &    -792 &         4a \\
       NGC 7311 &   2005kc &            &  72.34 & -43.41 &   4533 &   4612 &    4351 &  34.23 &  143 &     291 &       4b,5 \\
      UGC 12133 &   1998eg &            &  76.48 & -42.06 &   7423 &   7497 &    7238 &  35.46 &  146 &     -56 &      1,2,3 \\
       NGC 7329 &   2006bh &            & 320.97 & -45.79 &   3252 &   3058 & \nodata &  33.63 &   71 &    -228 &         4b \\
  Anon J2241-00 &   2006py &            &  68.44 & -48.79 &  17358 &  17372 &   17186 &  37.14 &  137 &    1333 &         4b \\
      UGC 12158 &   2004ef &            &  85.92 & -33.43 &   9289 &   9413 &    9096 &  35.80 &  157 &     606 &         4b \\
        IC 5270 &    1993L &            &   5.94 & -64.39 &   1983 &   1915 & \nodata &  32.03 &  101 &     335 &        1,3 \\
       NGC 7448 &   1997dt &            &  87.57 & -39.12 &   2194 &   2338 & \nodata &  32.99 &  152 &    -115 &        1,3 \\
   Anon 2304-37 &   1992aq &            &   1.78 & -65.31 &  30279 &  30107 &   30202 &  38.60 &  100 &    -285 &      1,2,3 \\
  MCG +05-54-41 &   2006en &            &  98.23 & -27.74 &   9575 &   9733 &    9387 &  36.00 &  164 &      97 &          5 \\
       NGC 7541 &   1998dh &            &  82.84 & -50.65 &   2689 &   2776 & \nodata &  32.94 &  140 &     378 &      1,2,3 \\
     ESO 291-11 &   1992bl &            & 344.15 & -63.93 &  12891 &  12701 &   12839 &  36.64 &   92 &    -138 &      1,2,3 \\
        Pegasus &  \nodata &    Pegasus &  87.78 & -48.38 &   3554 &   3633 &    3381 &  33.96 &  143 &    -188 &          6 \\
       NGC 7634 &    1972J &    Pegasus &  88.69 & -47.93 &   3225 &   3317 & \nodata &  33.53 &  144 &     177 &          1 \\
       NGC 7678 &   2002dp &            &  98.88 & -36.55 &   3489 &   3642 &    3308 &  33.65 &  155 &     325 &          5 \\
          A2634 &  \nodata &      A2634 & 103.45 & -33.06 &   9409 &   9546 &    9230 &  35.72 &  157 &    1051 &          6 \\
   Anon 2340+26 &   1997dg &      A2634 & 103.62 & -33.98 &  10193 &  10321 &   10014 &  36.22 &  156 &    -317 &      1,2,3 \\
          A4038 &  \nodata &      A4038 &  25.19 & -75.84 &   8994 &   8864 &    8904 &  35.82 &  106 &     -22 &          6 \\
     ESO 471-27 &   1993ah &      A4038 &  25.87 & -76.77 &   8803 &   8674 &    8714 &  35.70 &  106 &     256 &        1,3 \\
       NGC 7780 &   2001da &            &  99.20 & -52.06 &   5155 &   5214 &    4995 &  34.37 &  140 &     608 &          5 \\
     ESO 538-13 &   2005iq &            &  64.83 & -75.21 &  10206 &  10110 &   10097 &  36.15 &  114 &    -199 &       4a,b \\
% ***********************************************
\enddata
% ***********************************************

% ***********************************************
\end{deluxetable}
% ********************************************************

\clearpage

% ********************************************************
%  Table 2: Mean Random Modulus Differences Between Different Sources
% ********************************************************
\begin{deluxetable}{ccccccccc}
\tablewidth{0pt}
\tabletypesize{\scriptsize}
\tablecaption{\sc Mean Random (Absolute) Modulus Differences Between 
                  Different Sources\label{tab:02}}
% ***********************************************
\tablehead{
% ***********************************************
                            &
 \colhead{WAN-REI}          &  
 \colhead{JHA-REI}          &
 \colhead{WOO-REI}          & 
 \colhead{WAN-JHA}          &
 \colhead{WAN-WOO}          &
 \colhead{JHA-WOO}          &       
 \colhead{FRE-WOO}          & 
 \colhead{SNe-cl.}            
\\
                            &
 \colhead{(1)}              &  
 \colhead{(2)}              &
 \colhead{(3)}              & 
 \colhead{(4)}              &
 \colhead{(5)}              &
 \colhead{(6)}              &
 \colhead{(7)}              &
 \colhead{(8)}              
}
% ***********************************************
\startdata
% ***********************************************
$\sigma_{(m\!-\!M)}$: & 0.09 & 0.14 & 0.16 & 0.12 & 0.20 & 0.20 & 0.14 & 0.18 \\
N:                    &   81 &  101 &    7 &   88 &   15 &   15 & 6    &   14 \\
% ***********************************************
\enddata
% ***********************************************
% ***********************************************
\end{deluxetable}
% ********************************************************

% ********************************************************
%  Table 3: Properties of the Hubble Diagrams from Different Sources
% ********************************************************
\begin{deluxetable}{lcccccr}
\tablewidth{0pt}
\tabletypesize{\footnotesize}
\tablecaption{\sc Properties of the Hubble Diagrams from 
                  Different Sources\label{tab:03}}
% ***********************************************
\tablehead{
% ***********************************************
 \colhead{Sample}               &  
 \colhead{N}                    &
 \colhead{slope $a$}            & 
 \colhead{error\ $\epsilon(a)$} &
 \colhead{$\sigma_{(m\!-\!M)}$}   
\\
 \colhead{(1)}                  &  
 \colhead{(2)}                  &
 \colhead{(3)}                  & 
 \colhead{(4)}                  &
 \colhead{(5)}                
}
% ***********************************************
\startdata
% ***********************************************
 Reindl              &  62 & 0.188 & 0.004 & 0.16 \\ 
 Wang                &  55 & 0.190 & 0.003 & 0.13 \\
 Jha                 &  72 & 0.183 & 0.003 & 0.18 \\ 
 Wood-Vasey/Freedman &  45 & 0.189 & 0.005 & 0.16 \\ 
 Hicken              &  88 & 0.194 & 0.004 & 0.16 \\
 Masters             &  27 & 0.208 & 0.007 & 0.14 \\
 J{\o}rgensen        &  10 & 0.179 & 0.014 & 0.18 \\
\tableline\\[-9pt]
all                  & 218 & 0.193 & 0.002 & 0.15 \\
% ***********************************************
\enddata
% ***********************************************
%
% ***********************************************
\end{deluxetable}
% ********************************************************

\clearpage

% ********************************************************
%  Table 4: Delta(m-M) for different q_0
% ********************************************************
\begin{deluxetable}{cccccccccc}
\tablewidth{0pt}
\tabletypesize{\footnotesize}
\tablecaption{{\sc Modulus Differences} $\Delta(m-M)_{\rm model}$
  {\sc between Different World Models and the Model with}
  $\Omega_{\rm M}=0.3$, $\Omega_{\Lambda}=0.7$ ($q_{0}=-0.55)$ 
  {\sc in Function of} $z$\label{tab:q0}}
% ***********************************************
\tablehead{
% ***********************************************
 & & \multicolumn{7}{c}{$\Delta(m-M)_{\rm model}$} \\
 \colhead{$z$}                    &  
 \colhead{$\log z$}               &
 \colhead{        }               &
 \colhead{$\Omega_{\Lambda}=0.4$} &
 \colhead{$\Omega_{\Lambda}=0.9$} &
 \colhead{        }               &
 \colhead{$q_{0}= 0.00$}          & 
 \colhead{$q_{0}=-0.20$}          &
 \colhead{$q_{0}=-0.80$}          &
 \colhead{$q_{0}=-1.00$}        
\\
 \colhead{(1)}                    &  
 \colhead{(2)}                    &
 \colhead{}                       &
 \colhead{(3)}                    & 
 \colhead{(4)}                    &
 \colhead{}                       &
 \colhead{(5)}                    & 
 \colhead{(6)}                    &
 \colhead{(7)}                    &
 \colhead{(8)}              
}
% ***********************************************
\startdata
% ***********************************************
 $0.01$ & $-2.000$ & & $ 0.00$ & $-0.00$ & & $ 0.01$ & $ 0.00$ & $-0.00$ & $-0.00$ \\
 $0.02$ & $-1.699$ & & $ 0.01$ & $-0.01$ & & $ 0.01$ & $ 0.01$ & $-0.01$ & $-0.01$ \\
 $0.03$ & $-1.523$ & & $ 0.01$ & $-0.01$ & & $ 0.02$ & $ 0.01$ & $-0.01$ & $-0.01$ \\
 $0.04$ & $-1.398$ & & $ 0.02$ & $-0.01$ & & $ 0.02$ & $ 0.02$ & $-0.01$ & $-0.02$ \\
 $0.05$ & $-1.301$ & & $ 0.02$ & $-0.02$ & & $ 0.03$ & $ 0.02$ & $-0.01$ & $-0.02$ \\
 $0.06$ & $-1.222$ & & $ 0.03$ & $-0.02$ & & $ 0.04$ & $ 0.02$ & $-0.02$ & $-0.03$ \\
 $0.07$ & $-1.155$ & & $ 0.03$ & $-0.02$ & & $ 0.04$ & $ 0.03$ & $-0.02$ & $-0.03$ \\
 $0.08$ & $-1.097$ & & $ 0.04$ & $-0.03$ & & $ 0.05$ & $ 0.03$ & $-0.02$ & $-0.04$ \\
 $0.09$ & $-1.046$ & & $ 0.04$ & $-0.03$ & & $ 0.05$ & $ 0.03$ & $-0.02$ & $-0.04$ \\
 $0.10$ & $-1.000$ & & $ 0.05$ & $-0.03$ & & $ 0.06$ & $ 0.04$ & $-0.03$ & $-0.05$ \\
% ***********************************************
\enddata
% ***********************************************
%
% ***********************************************
\end{deluxetable}
% ********************************************************

% ********************************************************
%  Table 5: Convergence depth of the Local Supercluster
% ********************************************************
\begin{deluxetable}{llll}
\tablewidth{0pt}
\tabletypesize{\footnotesize}
\tablecaption{{\sc Convergence Depth of the Local Supercluster}.\label{tab:LSconvergence}}
% ***********************************************
\tablehead{
% ***********************************************
 \colhead{Author(s)}        &  
 \colhead{Method}           &
 \colhead{$N$}              &
 \colhead{convergence}      
\\
 \colhead{}                 &  
 \colhead{}                 &
 \colhead{}                 &
 \colhead{($\kms$)}      
% \\
%  \colhead{(1)}              &  
%  \colhead{(2)}              &
%  \colhead{(3)}              &
%  \colhead{(4)}              
}
% ***********************************************
\startdata
% ***********************************************
\citealt{Aaronson:etal:86}   & clusters (21cm)   & 10             & $<10,000$     \\
\citealt{Lilje:etal:86}      & 21cm              & field galaxies & $\sim4000$    \\
\citealt{Jerjen:Tammann:93}  & clusters          & 15             & $<6400$       \\
\citealt{FST:94}             & line widths       & field galaxies & $4000$        \\
\citealt{Riess:etal:95}      & SNe\,Ia           & 13             & $<7000$       \\
\citealt{Giovanelli:etal:98} & clusters (21cm)   & 24             & $<6000$       \\
\citealt{Dale:etal:99}       & clusters (21cm)   & 52             & $4000$-$6000$ \\
\citealt{Hoffman:etal:01}    & various           & field galaxies & $<6000$       \\
\citealt{Watkins:etal:09}    & various           & field galaxies & $5000$        \\
present paper                & SNe\,Ia, clusters & 170            & $3500\pm300$  \\
% ***********************************************
\enddata
% ***********************************************
%
% ***********************************************
\end{deluxetable}
% ********************************************************

% ********************************************************
%  Table 6: One-dimensional bulk velocities of (super-)clusters
% ********************************************************
\begin{deluxetable}{lccccr}
\tablewidth{0pt}
\tabletypesize{\footnotesize}
\tablecaption{{\sc One-Dimensional Bulk Velocities of (Super-) Clusters and
  Groups with} $3300<v_{\rm CMB}<10,000\kms$\label{tab:cluster}}
% ***********************************************
\tablehead{
% ***********************************************
 \colhead{cluster}                      &  
 \colhead{$\mu^{0}$}                    &
 \colhead{$N$}                          &
 \colhead{$v_{\rm CMB}$}                &  
 \colhead{$v_{\rm model}$}              &    
 \colhead{$v_{\rm CMB}-v_{\rm model}$}      
\\
 \colhead{(1)}                          &  
 \colhead{(2)}                          &  
 \colhead{(3)}                          &  
 \colhead{(4)}                          &  
 \colhead{(5)}                          &  
 \colhead{(6)}                               
}
% ***********************************************
\startdata
% ***********************************************
          A2806 &  35.47 &   1 &   8310 &   7588 & $ 722\pm623$ \\
         N383Gr &  34.54 &   2 &   4957 &   4977 & $ -20\pm248$ \\
          A2877 &  35.46 &   1 &   7396 &   7554 & $-158\pm555$ \\
         N507Gr &  34.31 &   2 &   4802 &   4482 & $ 319\pm240$ \\
          A0194 &  34.60 &   2 &   5302 &   5115 & $ 187\pm265$ \\
          A0262 &  34.44 &   1 &   4770 &   4756 & $  14\pm358$ \\
          A0397 &  35.87 &   2 &   9749 &   9089 & $ 660\pm487$ \\
          A0400 &  35.19 &   1 &   7276 &   6685 & $ 591\pm546$ \\
          A0426 &  34.68 &   2 &   5745 &   5304 & $ 441\pm287$ \\
          A0496 &  36.03 &   1 &   9928 &   9767 & $ 160\pm745$ \\
          A0539 &  35.95 &   1 &   8580 &   9422 & $-842\pm643$ \\
          A0569 &  34.92 &   3 &   6053 &   5915 & $ 138\pm303$ \\
          A0634 &  35.72 &   1 &   7957 &   8495 & $-538\pm597$ \\
         Cancer &  34.51 &   2 &   4623 &   4910 & $-286\pm231$ \\
          A0779 &  35.52 &   1 &   6840 &   7761 & $-921\pm513$ \\
          Hydra &  34.23 &   2 &   3973 &   4322 & $-349\pm199$ \\
          AS639 &  34.89 &   1 &   6512 &   5835 & $ 677\pm488$ \\
          A1367 &  34.98 &   2 &   6707 &   6078 & $ 629\pm335$ \\
           Coma &  35.49 &   3 &   6982 &   7657 & $-675\pm349$ \\
          A3574 &  34.42 &   3 &   4913 &   4713 & $ 200\pm246$ \\
          AS753 &  34.27 &   1 &   4306 &   4401 & $ -95\pm323$ \\
          A2199 &  35.81 &   2 &   8949 &   8847 & $ 103\pm447$ \\
         PavoII &  34.28 &   2 &   4189 &   4422 & $-233\pm209$ \\
          Grm15 &  34.22 &   2 &   4269 &   4302 & $ -34\pm213$ \\
          PavoI &  33.98 &   1 &   4138 &   3857 & $ 281\pm310$ \\
        Pegasus &  33.75 &   2 &   3381 &   3472 & $ -91\pm169$ \\
          A2634 &  35.97 &   2 &   9230 &   9507 & $-278\pm461$ \\
          A4038 &  35.76 &   2 &   8904 &   8649 & $ 255\pm445$ \\
% ***********************************************
\enddata
% ***********************************************
%
% ***********************************************
\end{deluxetable}
% ********************************************************

% ******************************************************************
% ***********              Figures                       ***********
% ******************************************************************
\clearpage

% ********************************************************
% Figure 1: Hubble diagrams of (1) - (8)
% ********************************************************
% ********************************************************
\clearpage
% ********************************************************

% ********************************************************
%  Figure 2: Combined Hubble diagram
% ********************************************************
% ********************************************************

% ********************************************************
% Figure 3: Hubble diagram - Dn-sigma
% ********************************************************
% ********************************************************

% ********************************************************
% Figure 4: Hubble diagram - SBF
% ********************************************************
% ********************************************************

% ********************************************************
% Figure 5: Hubble diagram - PNLF
% ********************************************************
% ********************************************************

% ********************************************************
%  Figure 6: Deviations Delta log v from the adopted Hubble line
% ********************************************************
% ********************************************************

% ********************************************************
%  Figure 7: Modulus differences Delta(m-M) vs. log z
% ********************************************************
% ********************************************************

% ********************************************************
% Figure 8: Hypothetical Hubble diagram - Apex vs. Antapex
% ********************************************************
% ********************************************************

% ********************************************************
%  Figure 9: Velocity residuals
% ********************************************************
% ********************************************************

% ********************************************************
% Figure 10: Velocity residuals (4x2)
% ********************************************************
% ********************************************************

% ********************************************************
%  Figure 11: Aitoff Diagram
% ********************************************************
% ********************************************************

% ********************************************************
%  Figure 12: Flow diagram
% ********************************************************
% ********************************************************

% ******************************************************************
\end{document}